\documentclass[useAMS,usenatbib]{mnras}
\usepackage[dvips]{graphicx}
\usepackage[english]{babel}
\usepackage{amsmath}
\usepackage{amssymb}
\usepackage{textcomp}
\usepackage{natbib}

\title[The growth of massive galaxies]{Galaxy growth from redshift 5 to 0 at fixed comoving number density}
\author[F. van de Voort]{Freeke van de Voort$^{1,2}$\thanks{E-mail: freeke@berkeley.edu} \\
$^{1}$Department of Astronomy and Theoretical Astrophysics Center, University of California, Berkeley, CA 94720-3411, USA \\
$^{2}$Academia Sinica Institute of Astronomy and Astrophysics, P.O. Box 23-141, Taipei 10617, Taiwan}

\begin{document}

\date{Accepted July 5, 2016. Received June 3, 2016; in original form June 19, 2015}

\pagerange{\pageref{firstpage}--\pageref{lastpage}} \pubyear{2015}

\maketitle

\label{firstpage}

\begin{abstract}

Studying the average properties of galaxies at a fixed comoving number density over a wide redshift range has become a popular observational method, because it may trace the evolution of galaxies statistically. We test this method by comparing the evolution of galaxies at fixed number density and by following individual galaxies through cosmic time ($z=0-5$) in cosmological, hydrodynamical simulations from OWLS. Comparing progenitors, descendants, and galaxies selected at fixed number density at each redshift, we find differences of up to a factor of three for galaxy and interstellar medium (ISM) masses. The difference is somewhat larger for black hole masses. The scatter in ISM mass increases significantly towards low redshift with all selection techniques. We use the fixed number density technique to study the assembly of dark matter, gas, stars, and black holes and the evolution in accretion and star formation rates. We find three different regimes for massive galaxies, consistent with observations: at high redshift the gas accretion rate dominates, at intermediate redshifts the star formation rate is the highest, and at low redshift galaxies grow mostly through mergers. Quiescent galaxies have much lower ISM masses (by definition) and much higher black hole masses, but the stellar and halo masses are fairly similar. Without active galactic nucleus (AGN) feedback, massive galaxies are dominated by star formation down to $z=0$ and most of their stellar mass growth occurs in the centre. With AGN feedback, stellar mass is only added to the outskirts of galaxies by mergers and they grow inside-out.

\end{abstract}

\begin{keywords}
galaxies: evolution -- galaxies: formation -- galaxies: haloes -- galaxies: statistics -- intergalactic medium -- methods: numerical
\end{keywords}

\section{Introduction}

Galaxies grow through accretion of diffuse and filamentary gas and through mergers with a wide range of mass ratios. This growth is thought to be regulated by feedback from star formation and active galactic nuclei (AGN) \citep[see e.g.][]{Mo2010}. The material expelled in galactic winds almost balances the gas brought in by accretion. Without outflows, cosmological hydrodynamical simulations overproduce the stellar mass formed in both low-mass and high-mass haloes as well as the baryon fraction in groups and clusters \citep[e.g.][]{White1991, Benson2003, McCarthy2010}. 

Cosmological simulations try to include the relevant physical processes for galaxy formation, but need to employ simplified subgrid recipes to model stellar and black hole feedback. In recent years, huge progress has been made in understanding the effects of these subgrid models and in capturing important properties of observed galaxies and their haloes \citep[e.g.][]{Schaye2010, Dave2011, Vogelsberger2014, Genel2014, Schaye2015, Crain2015}. The growth of structure is followed from the early Universe until the present day and simulations are therefore a powerful tool to study galaxy formation dynamically. However, there are also limitations to how accurately simulations can produce realistic galaxies and it is therefore vital to directly compare them to observations. Doing so enables us to both validate or invalidate aspects of simulations and helps us to interpret observations of galaxies. 

Galaxies are observed and studied at a large range of redshifts ($z=0-10$) and are found to have very different properties at low ($z=0-1$) and high redshift ($z=2-6$). For example, galaxies at high redshift are, on average, more clumpy, more compact, and have higher star formation rates than equally-massive galaxies at low redshift \citep[e.g.][]{Damen2009, Genzel2011, Szomoru2012}. An important question is how these different populations are connected.

To understand how galaxies at high redshift grow into their present-day counterparts, one cannot compare galaxies at different epochs at fixed mass. Rather, the galaxies need to be linked while taking into account their mass growth. Using hydrodynamical simulations or semi-analytic models, it is straightforward to find progenitors and descendants of galaxies and thus calculate the rate at which they grow \citep[e.g.][]{Voort2011a}. Observationally, however, it is hard to connect populations of galaxies at different epochs.

One purely observational approach is to look at the galaxy population at fixed comoving number density. By assuming that the brightest galaxies at high redshift evolve into the brightest galaxies at low redshift, one can infer the rate at which the interstellar medium (ISM) mass and stellar mass grow \citep[e.g.][]{Dokkum2010, Papovich2011, Ownsworth2014} and one can study how the structural properties of galaxies change with time \citep[e.g.][]{Brammer2011, Dokkum2013, Patel2013a, Muzzin2013, Papovich2015}. Some studies have been performed using only star-forming galaxies \citep{Patel2013b}.

A hybrid technique, using both observations and simulations, is the abundance matching method in which observed galaxies are linked to simulated dark matter haloes by matching their spatial density \citep[e.g.][]{Conroy2009, Moster2013, Behroozi2013a}. In this method the most massive galaxies are assumed to reside in the most massive haloes, somewhat less massive galaxies inside somewhat less massive haloes, etcetera. Additional information, such as star formation rates, can be used to break degeneracies. By using merger trees, the growth of haloes can be computed from the dark matter simulations directly, from which the growth of the corresponding galaxies can be derived. Recently, a number of observational studies have used results from abundance matching to correct for mergers by decreasing the number density with decreasing redshift \citep{Marchesini2014, Ownsworth2014, Papovich2015}.

Abundance matching, semi-analytic models, and cosmological simulations provide ways to test how well galaxies can be traced by number density selections \citep{Leja2013, Behroozi2013b, Mundy2015}. Vice versa, using number densities, we can directly compare evolving galaxy properties, like star formation rates and quiescent fractions, in simulations to those in observations.

Here, we use cosmological, hydrodynamical simulations to compare the growth of galaxies and haloes at fixed comoving number density with the mass evolution of their descendants or progenitors. We show that the difference is reasonably small, which lends support to the use of this method. For the first time we show the evolution of each individual component: halo mass, ISM mass, stellar mass, and supermassive black hole mass. We also include accretion rates and star formation rates and find that there are three regimes of growth for massive galaxies. At high redshift, gas accretion dominates. At intermediate redshift, star formation dominates. At low redshift, mergers dominate. We investigate the importance of AGN feedback and compare simulations with and without AGN feedback. We find that without AGN feedback at low redshift star formation dominates strongly down to $z=0$, resulting in more massive galaxies. With AGN feedback the galaxies become quenched and grow mainly in their outer parts (or `inside-out'). 

This paper is organized as follows. We describe the simulations used in Section~\ref{sec:sim}.  In Section~\ref{sec:method} the methods used to connect galaxies at various redshifts are described. Section~\ref{sec:results} shows the mass growth and accretion rates from $z=5$ to $z=0$ for our fiducial simulation. In Section~\ref{sec:compare} we compare the results from Section~\ref{sec:results} to a simulation without AGN feedback. We discuss how our simulated galaxies relate to certain observations in Section~\ref{sec:obs}. Our conclusions can be found in Section~\ref{sec:concl}.

\section{Simulations} \label{sec:sim}

We use a modified version of \textsc{gadget-3} \citep[last described in][]{Springel2005}, a smoothed particle hydrodynamics (SPH) code that uses the entropy formulation of SPH \citep{Springel2002}, which conserves both energy and entropy where appropriate. This work is part of the OverWhelmingly Large Simulations (OWLS) project \citep{Schaye2010}, which consists of a large number of cosmological simulations with varying (subgrid) physics. For our main results we make use of the AGN model (\emph{``AGN''}). The model is fully described in \citet{Schaye2010} and we will summarize its main properties here. For comparison and resolution tests we also use the reference model (\emph{``REF''}), which is identical to the AGN model except for the omission of AGN feedback. 

The simulations assume a $\Lambda$CDM cosmology with parameters $\Omega_\mathrm{m} = 1 - \Omega_\Lambda = 0.238$, $\Omega_\mathrm{b} = 0.0418$, $h = 0.73$, $\sigma_8 = 0.74$, $n = 0.951$. These values are consistent with the WMAP 9-year data \citep{Hinshaw2013}. The only significant discrepancy is in $\sigma_8$, which is 10 per cent lower than the value favoured by the WMAP 9-year data.

A cubic volume with periodic boundary conditions is defined, within which the mass is distributed over $512^3$ dark matter and as many gas particles. The volume of the simulations used in this work is (100~comoving~$h^{-1}$Mpc)$^3$. The effect of having a finite simulation volume for the present work is described in Appendix~\ref{sec:box} and found to be negligible. The (initial) particle masses for baryons and dark matter are $1.2\times10^8$~M$_\odot$ and $5.6\times10^8$~M$_\odot$, respectively. 
The gravitational softening length is 7.8~comoving~$h^{-1}$kpc, i.e.\ 1/25 of the mean dark matter particle separation, but is limited to a maximum value of 2~proper~$h^{-1}$kpc, which is reached at $z=2.91$.

The initial conditions are set up as follows. Particles are placed in the simulation box in an initially glass-like state, as described in \citet{White1994}. The particles are then displaced according to a theoretical power spectrum, which is generated using \textsc{cmbfast} \citep[][version 4.1]{Seljak1996}. The particles are subsequently evolved down to redshift z = 127 using the Zel'Dovich approximation \citep{Zeldovich1970} at which point the simulation is started.

Star formation is modelled according to the recipe of \citet{Schaye2008}. A polytropic equation of state $P_\mathrm{tot}\propto\rho_\mathrm{gas}^{4/3}$ is imposed for densities exceeding $n_\mathrm{H}^\star=0.1$~cm$^{-3}$, where $P_\mathrm{tot}$ is the total pressure and $\rho_\mathrm{gas}$ the density of the gas. Gas particles with proper densities $n_\mathrm{H}\ge0.1$~cm$^{-3}$ and temperatures $T\le10^5$~K are put on this equation of state and have a finite probability to be converted into star particles. The star formation rate (SFR) per unit mass depends on the gas pressure and reproduces the observed Kennicutt-Schmidt law \citep{Kennicutt1998} by construction. We identify the star-forming gas on the equation of state as the ISM of the galaxy.

Feedback from star formation is implemented using the prescription of \citet{Vecchia2008}. About 40 per cent of the energy released by type II supernovae is injected locally in kinetic form, while the rest of the energy is assumed to be lost radiatively. The initial wind velocity is 600~km~s$^{-1}$ and the initial mass loading is two, meaning that, on average, a newly formed star particle kicks twice its own mass in neighbouring gas particles. These values were chosen to roughly reproduce observations of the global SFR density.

The abundances of eleven elements released by massive stars and intermediate mass stars are followed as described in \citet{Wiersma2009b}. We assume the stellar initial mass function (IMF) of \citet{Chabrier2003}, ranging from 0.1 to 100~M$_\odot$. As described in \citet{Wiersma2009a}, radiative cooling and heating are computed element by element in the presence of the cosmic microwave background radiation and the \citet{Haardt2001} model for the UV/X-ray background from galaxies and quasars, assuming the gas to be optically thin and in (photo-)ionization equilibrium.

The model for the formation of and feedback from AGN is fully described and tested in \citet{Booth2009}. It is a modified version of the model introduced by \citet{Springeletal2005}. A seed mass black hole of $10^5$~M$_\odot$ is placed in every resolved halo. These black holes grow by accretion of gas, after which energy is injected into the surrounding medium, and by mergers. The accretion rate onto the black hole equals the so-called Bondi-Hoyle accretion rate \citep{Bondi1944} if the gas density is low ($n_{\rm H} < 10^{-1}\,{\rm cm}^{-3}$). However, in dense, star-forming gas the accretion rate would be severely underestimated because the simulations do not include a cold, interstellar gas phase and because the Jeans scales are unresolved. The Bondi-Hoyle accretion rate is multiplied by an efficiency parameter $\alpha=(n_\mathrm{H}/n^\star_\mathrm{H})^\beta$, where $\beta=2$.  Note, however, that our results are insensitive to the choice for $\beta$ as long as it is chosen to be sufficiently large (see \citealt{Booth2009}). 

In our fiducial simulation (with AGN feedback), a fraction of 1.5 per cent of the rest-mass energy of the gas accreted onto the black hole is injected into the surrounding medium in the form of heat, by increasing the temperature of at least one neighbouring gas particle by at least $10^8$~K. The minimum temperature increase ensures that the feedback is effective, because the radiative cooling time of the heated gas is sufficiently long, and results in fast outflows. When AGN feedback is included, the SFR is reduced for haloes with $M_\mathrm{halo}\gtrsim10^{12}$~M$_\odot$ \citep{Booth2009}. The AGN simulation reproduces the observed mass density in black holes at $z=0$ and the black hole scaling relations \citep{Booth2009} and their evolution \citep{Booth2011} as well as the observed optical and X-ray properties, gas fractions, SFRs, stellar age distributions and the thermodynamic profiles of groups of galaxies \citep{McCarthy2010}. 

Since the simulation without AGN feedback approximately matches the global SFR density, adding AGN feedback reduces it to somewhat below the observed value \citep{Schaye2010}. This affects the normalization of the stellar mass, which is somewhat low. OWLS were not tuned to match the galaxy stellar mass function, which is therefore not reproduced in exact detail. The fraction of massive galaxies that is quenched is in agreement with observations, as shown in this paper. The fact that the simulations match many observations with minimal tuning, lends credence to the idea that they are following the main processes important for galaxy evolution, at least at the more massive end \citep[e.g.]{Crain2010a, McCarthy2010, McCarthy2012, Booth2011, Voort2012, Haas2013}.  

The simulation data is saved at discrete output redshifts with interval $\Delta z= 0.125$ at $0\le z\le 0.5$, $\Delta z = 0.25$ at $0.5 <z \le 4$
and $\Delta z = 0.5$ at $4 < z \le 5$. This is the time-resolution used for tracing the galaxy population and determining their properties.

We performed a resolution study with the reference model, which does not include feedback from AGN. This simulation was performed with an eight (two) times higher mass (spatial) resolution in a volume that was eight times smaller than in our fiducial case. The simulations are in good agreement with each other, so we conclude that our results do not strongly depend on resolution.

\section{Method} \label{sec:method}

In this paper, we study central galaxies at cumulative comoving number densities of $n(>M)=2\times10^{-5}$, $2\times10^{-4}$, and $2\times10^{-3}$~Mpc$^{-3}$, where $n(>M)$ is the number density of haloes with mass larger than $M$. This corresponds to $N=51$, 514, and 5141 galaxies, respectively, in our simulation volume of (100~$h^{-1}$Mpc)$^3$ = 2.57$\times10^6$~comoving~Mpc$^3$. At $z=0$ these number densities correspond to median halo masses of $10^{14.9}$, $10^{13}$, and $10^{12.1}$~M$_\odot$ and median stellar masses of $10^{11.5}$, $10^{10.8}$, and $10^{10.1}$~M$_\odot$, respectively, in our fiducial simulation. We choose $n(>M)=2\times10^{-4}$~Mpc$^{-3}$ to match the number density used in \citet{Papovich2011} and a factor of ten larger and smaller to bracket other common choices. I also probes a very interesting mass range, where the vast majority of galaxies change from being star-forming at $z=5$ to being quenched at $z=0$. We rank central galaxies according to their halo mass and disregard satellite galaxies. We then select the most massive haloes until the chosen number density is reached.

Because we wish to study the evolution of galaxies from $z=5$ to the present, we need to link them across cosmic time. We follow three different methods. The first one (`nrdens') is where we select the central galaxies of the $N$ most massive haloes at each redshift, corresponding to a fixed number density. For the second method (`prog') we select the central galaxies of the $N$ most massive haloes at $z=0$. These galaxies are then traced back to higher redshift by identifying their most massive progenitors at each output redshift. The third method (`desc') selects the central galaxies of the $N$ most massive haloes at $z=5$. These galaxies are traced down to $z=0$ by finding their descendants. The identification of progenitors and descendants is described in Section \ref{sec:id}. We also show results for a combined method (`prog+desc') where we select the $N$ most massive galaxies at $z=2$ and trace progenitors to $z=5$ and descendants to $z=0$.

We do not expect selecting on halo mass instead of stellar mass to bias the results, because stellar mass increases with halo mass and we are interested in studying the galaxy population statistically. We confirmed this by redoing our analysis ranking all galaxies according to their stellar masses, as shown and described in Appendix~\ref{sec:star}. The differences are small and our conclusions remain unchanged. The only significant difference is that the scatter in stellar masses is somewhat reduced when selecting on stellar mass. 

Since the majority of massive galaxies are central galaxies, including or excluding satellites has only a small effect. We choose to study central galaxies alone, in order to also investigate their halo properties at the virial radius, which is not well-defined for satellites. We redid our analysis including satellite galaxies, also shown in Appendix~\ref{sec:star} and find that most of our results do not change quantitatively. The exception is the fraction of descendants recovered, as described later in Section~\ref{sec:id}. Differences are somewhat larger for higher number densities, since at low stellar mass more galaxies are satellites. 

Previous work also includes studies of the evolution of galaxies in number density bins,  i.e.\ all galaxies with mass similar to the lowest mass in the cumulative number density selection \citep[e.g.][]{Patel2013a, Dokkum2013}. Because in this paper we calculate the median properties of the galaxies in the sample and because there are many more low-mass galaxies, results are weighted towards the low-mass end. Median masses are somewhat lower when using number density bins, but the trends remain the same (see also Appendix~\ref{sec:star}).

\subsection{Identifying haloes and galaxies}

In order to identify haloes, and galaxies that reside within them, we first find dark matter haloes using a Friends-of-Friends (FoF) algorithm. If the separation between two dark matter particles is less than 20 per cent of the average dark matter particle separation, they are placed in the same group. The minimum number of dark matter particles in a FoF group is set to 25. Many of the smallest groups will not be gravitationally bound. Baryonic particles are placed in a group if their nearest dark matter neighbour is part of the group. We then use \textsc{subfind} \citep{Springel2001, Dolag2009} on the FoF output to find the gravitationally bound particles and to identify subhaloes. We use these subhaloes to test for the effect of including satellites. 

For the results shown in this paper we use the most bound particle of an FoF halo, as found by \textsc{subfind} as the halo centre. We then use a spherical overdensity criterion, considering all the particles in the simulation. We compute the virial radius, $R_\mathrm{vir}$, within which the average density agrees with the prediction of the top-hat spherical collapse model in a $\Lambda$CDM cosmology \citep{Bryan1998}. 

For each halo, we identify the ISM of the central galaxy with the star-forming (i.e.\ $n_\mathrm{H}>0.1$~cm$^{−3}$) gas particles in the main halo which are inside 15 per cent of $R_\mathrm{vir}$ \citep{Sales2010}. We use 15 per cent of the virial radius to exclude unresolved star-forming satellites in the outer halo. Similarly, we calculate the stellar mass of a galaxy by taking into account all the stars within $0.15R_\mathrm{vir}$. The difference with using the total stellar mass within the virial radius is small, on average 0.1~dex.

The simulations we use in this paper reproduce the halo mass function, but do not exactly reproduce the galaxy stellar mass function. The stellar masses shown should therefore be interpreted with some care. Galaxy mergers, scatter in the growth history of haloes and galaxies, and quenching of massive galaxies due to AGN feedback are all included. We have repeated our analysis for a simulation without AGN feedback (which overpredicts the masses of massive galaxies) and found similar differences when comparing different selection methods, so these results are robust to changes in feedback model. These simulations exhibit mass growth over time and scatter in properties that can be very different, which is shown and discussed in Section~\ref{sec:compare}. They are contrasted with observational studies of galaxy evolution at fixed number density in Section~\ref{sec:obs}, which enables easy comparison of trends with redshift. Based on this and previous work, we consider our fiducial simulation with AGN feedback the most realistic. \citet{Torrey2015} use a simulation that approaches the observed stellar mass function more closely and find trends with redshift and differences between progenitors, descendants, and galaxies at fixed number density that are in quantitative agreement with ours. The normalization of the masses is indeed somewhat different. Note that these authors select galaxies in number density bins, which also leads to small normalization differences.

\subsection{Identifying progenitors and descendants} \label{sec:id}

For a halo at $z = z_2$, we identify its progenitor at the previous output redshift $z_1 > z_2$ and its descendant at the next output redshift $z_3<z_2$. We determine which progenitor (descendant) candidate contains most of the halo's 25 most bound dark matter particles and refer to this halo as `the progenitor' (`the descendant'). If the fraction of the halo's 25 most bound particles that was not in any halo at $z_1$ is greater than the fraction that was part of the progenitor, then the progenitor is not resolved and we set its mass to zero. If two or more haloes contain the same number of those 25 particles, we select the one that contains the dark matter particle that is most bound to the original halo.

Progenitors cannot always be identified. If no progenitor can be found, because it is below the resolution limit of our simulation, or if the progenitor is displaced by more than 2~$h^{-1}$Mpc and thus probably misidentified, the mass of the progenitor is set to zero. This is only important for the lowest-mass galaxies, in the highest number density selection at $z\ge4$, and does not greatly affect our results. 

More importantly, the descendant is not necessarily unique. When two haloes merge, they have the same descendant. We discard duplicate haloes from our analysis. The number of haloes in our analysis of the descendants therefore decreases with decreasing redshift. Again, this effect is strongest for the highest number density selection. The reader should keep in mind that the number of descendant galaxies changes as a function of redshift.

\begin{figure}
\center
\includegraphics[scale=.5]{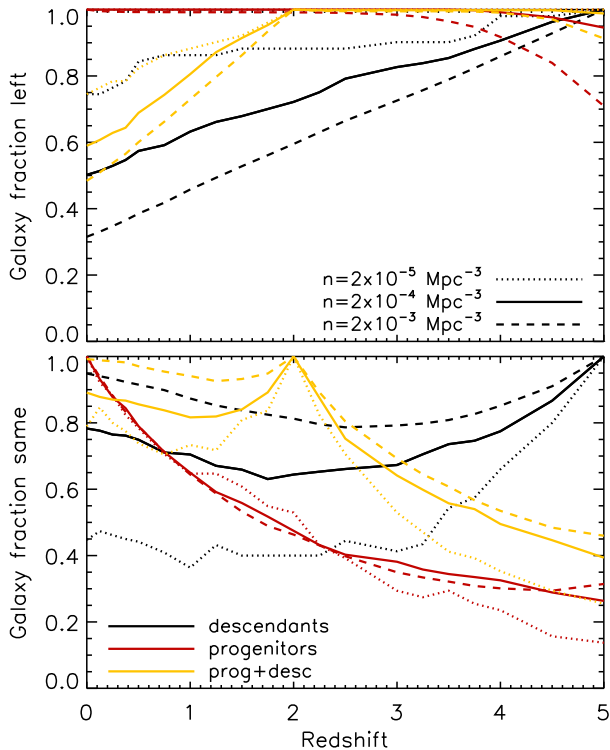}
\caption {\label{fig:leftsame} \textit{Top panel:} Fraction of galaxies left after tracing progenitors of $z=0$~selected galaxies (red curves), descendants of $z=5$~selected galaxies  (black curves), and progenitors and descendants of $z=2$~selected galaxies (`prog+desc', yellow curves) for different number densities used to select galaxies as indicated in the legend. The number of progenitors decreases significantly only for the highest number density selection at $z>4$ due to the limited resolution of the simulation. The number of descendants decreases strongly for all number density selections due to mergers and strongest for the highest number density selection. \textit{Bottom panel:} Fraction of individual galaxies in each sample that are also in the fixed number density sample selection. Except in the lowest number density bin, the majority of descendant galaxies are recovered by the number density selection. The majority of the progenitors, however, are not amongst the $N$ most massive galaxies after a redshift difference $\Delta z\gtrsim2$.}
\end{figure}

The top panel of Figure~\ref{fig:leftsame} shows the percentage of resolved progenitors and the percentage of unique descendants for our three number density selections at all redshifts. Black curves show the fraction of unique descendants of galaxies. Red curves show the fraction of resolved progenitors of $z=0$ selected galaxies. Yellow curves show the fraction of unique descendants and resolved progenitors of $z=2$ selected galaxies. We are able to trace progenitors out to high redshift. Even for the lowest mass (i.e.\ highest number density) sample we can still retrieve 98 (71) per cent of the $z=0$ selected galaxies at $z=3$ ($z=5$). For descendants the number of galaxies left is equal to one minus the fraction of mergers between galaxies in the sample. Mergers are important at each number density, but most severe at the highest number density. At $z=0$ we are left with 75, 50, and 32 per cent of the original number of $z=5$ selected galaxies. Although difficult to compare directly, because of differences in the selection of galaxies, this is qualitatively consistent with \citet{Behroozi2013b}. The fact that at higher number density, we find a lower number of galaxies left at low redshift does not mean that low-mass galaxies have higher halo merger rates, since we include a much wider range of galaxy masses at higher number density and thus allow for much larger merger ratios. 

The `merger fraction' is the quantity least robust to the inclusion or exclusion of satellites. When including satellites the trends are the same, but the fraction of galaxies left at $z=0$ for descendants of massive galaxies selected at $z=5$ is 82, 64, and 48 per cent for $n(>M)=2\times10^{-5}$, $2\times10^{-4}$, and $2\times10^{-3}$~Mpc$^{-3}$, respectively, and 92, 75, and 64 per cent for descendants of $z=2$~selected galaxies. This is higher than when only using central galaxies. 

We can also check how many galaxies in the descendant or progenitor sample are identical to one in the number density sample and thus still amongst the most massive. The fraction of galaxies which are also in the fixed number density selection is plotted in the bottom panel of Figure~\ref{fig:leftsame}. Although reducing the number of galaxies in the sample due to mergers, tracing descendants results in a large fraction (more than three quarters) of galaxies that are the same as in the pure number density selection, except for descendants of $z=5$ galaxies in our lowest number density sample of which only 45 per cent are amongst the most massive ($n(>M)=2\times10^{-5}$~Mpc$^{-3}$) at $z=0$. Additionally, there is no strong trend with redshift. Amongst progenitor galaxies, only 15-45 per cent are contained in our fixed number density selections at $z=5$ and this number decreases most steeply in the early phase of tracing progenitors. The $z=0$ ($z=2$) selected progenitor sample is at least 50 per cent complete out to $z\approx2$ ($z\approx4$). Vice versa, the $z=5$ ($z=2$) selected descendants are 50 per cent complete to $z=2.5$ ($z=0$).

When including satellites, the fraction of galaxies still in the sample that is the same as in the number density selected sample is 33, 56, and 72 per cent for descendants of $z=5$~selected galaxies and 60, 65, and 82 per cent for descendants of $z=2$~selected galaxies, for $n(>M)=2\times10^{-5}$, $2\times10^{-4}$, and $2\times10^{-3}$~Mpc$^{-3}$, respectively. This is significantly lower than when only using central galaxies. However, the fraction of resolved galaxies that are the same as in the number density selection (a combination of top and bottom panel of Figure~\ref{fig:leftsame}) is very similar.

\section{Mass evolution and accretion history} \label{sec:results} 

\subsection{Tracing methods}

\begin{figure}
\center
\includegraphics[scale=.5]{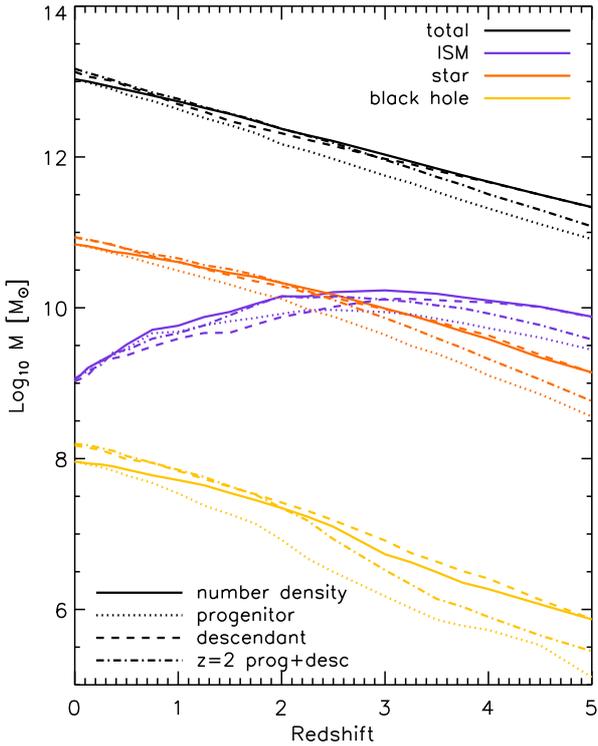}
\caption {\label{fig:nrprogdesc} Evolution of the median total halo mass (black curves), ISM mass (purple curves), stellar mass (orange curves) and black hole mass (yellow curves) using different methods to link high-redshift with low-redshift galaxies. The solid curves show median masses at fixed number density ($n(>M)=2\times10^{-4}$~Mpc$^{-3}$). Dotted (dashed) curves use $z=5$ ($z=0$) galaxies with the same number density, but trace their descendants (progenitors). Dot-dashed curves use $z=2$ galaxies with $n(>M)=2\times10^{-4}$~Mpc$^{-3}$ and trace progenitors to $z=5$ and descendants to $z=0$. The differences between selections increase to maximally 0.5~dex for dark matter, ISM, and stellar masses.}
\end{figure}

We first focus on the evolution of fixed number density populations, progenitors of massive galaxies selected at $z=0$, descendants of galaxies selected at $z=5$, and progenitors and descendants of galaxies selected at $z=2$. Figure \ref{fig:nrprogdesc} compares these four selections described in Section \ref{sec:method} to connect galaxies at different redshifts. It shows the median evolution of the total halo mass (black curves), ISM mass (purple curves), stellar mass (orange curves), and black hole mass (yellow curves) from $z=0$ to $z=5$. Solid curves show the mass evolution of galaxies at fixed number density $n(>M)=2\times10^{-4}$~Mpc$^{-3}$. Dotted (dashed) curves show results for galaxies selected by tracing the most massive progenitors (descendants) of haloes with $n(>M)=2\times10^{-4}$~Mpc$^{-3}$ at $z=0$ ($z=5$). Dot-dashed curves trace progenitors and descendants of $z=2$~selected galaxies. 

Tracing descendants of $z=5$~selected galaxies is remarkably similar to tracing galaxies at fixed number density, with the largest difference of 0.3~dex found for ISM masses around $z=2$. Note, however, that we take the median of fewer galaxies in the case of descendants, because some of the galaxies merged at higher redshift (see Figure~\ref{fig:leftsame}). The difference between fixed number densities and progenitors increases steadily towards higher redshift and reaches $0.4-0.7$~dex at $z=5$, with the smallest difference for total halo mass and the largest for black hole mass. At $z=2$ the different selections span about $0.2$~dex for dark matter, stellar, and ISM mass. 

Following the evolution of progenitors or descendants gives different answers to the question how fast galaxies grow, in agreement with \citet{Behroozi2013b, Mundy2015} and \citet{Torrey2015}. These differences are as large as the differences found when comparing to galaxy growth at fixed number densities. Physically, both tracing progenitors and descendants are equally valid ways to define galaxy growth. Therefore, given that tracing progenitors or descendants is inherently different, and given the many other uncertainties when determining galaxy masses, we conclude that tracing massive galaxies at fixed number density is a useful way to connect galaxies across time observationally as well as to directly compare observations to simulations.

\begin{figure}
\center
\includegraphics[scale=.5]{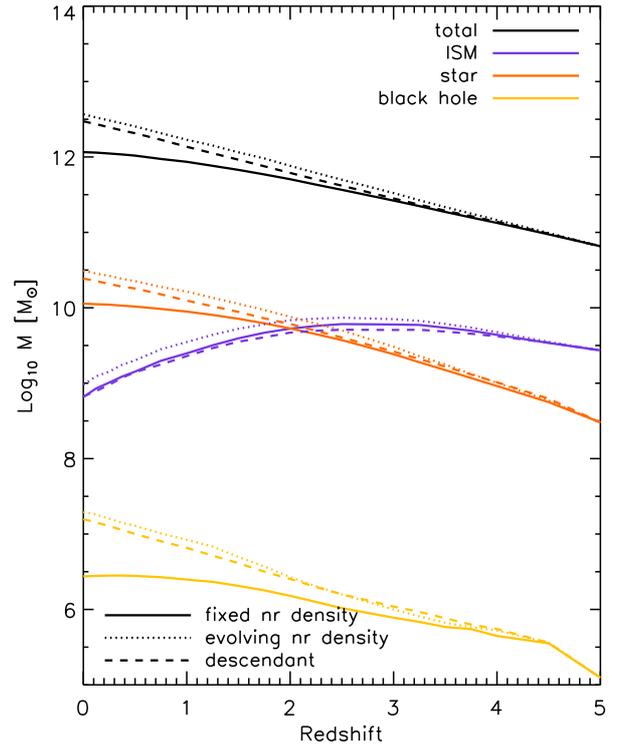}
\caption {\label{fig:nrevol5141} Evolution of the median total halo mass (black curves), ISM mass (purple curves), stellar mass (orange curves) and black hole mass (yellow curves) using different methods to link high-redshift with low-redshift galaxies. The solid curves show median masses at fixed number density ($n(>M)=2\times10^{-3}$~Mpc$^{-3}$, a factor of 10 higher than our fiducial value). Dotted curves use $z=5$ ($z=0$) galaxies with an evolving number density based on the number of mergers. Dashed curves curves use $z=5$ galaxies with the same number density, but trace their descendants down to $z=0$ and therefore include the same number of objects as the evolving number density selection. For $n(>M)=2\times10^{-3}$~Mpc$^{-3}$, an evolving number density traces descendants better than a fixed number density.}
\end{figure}
The differences between different tracing methods are similar for $n(>M)=2\times10^{-5}$~Mpc$^{-3}$, except that the descendant masses are lower than at fixed number density (by up to 0.1~dex for stellar mass and 0.5~dex for ISM mass), due to scatter in the growth rates and fewer mergers. For $n(>M)=2\times10^{-3}$~Mpc$^{-3}$ the different selection methods span about $0.3$~dex at $z=2$. The biggest difference, however, is seen at $z=0$, where the descendant halo, stellar and black hole masses are much larger than those at fixed number density. In this case, the scatter in growth rates have a much smaller effect than for lower number density selections, but the number of mergers within the sample is much larger. This can be corrected for by using an number density that evolves with redshift instead of a fixed number density, which has become a popular technique \citep[e.g.][]{Behroozi2013b, Marchesini2014, Papovich2015}. 

In Figure~\ref{fig:nrevol5141} we compare selecting galaxies at a fixed number density ($n(>M)=2\times10^{-3}$~Mpc$^{-3}$; solid curves) and at an evolving comoving number density (dotted curves) and selecting descendants of $z=5$ galaxies (dashed curves). This last selection takes into account the number of mergers occurring in the simulation when tracing descendants, as shown by the top panel of Figure~\ref{fig:leftsame}. The resulting median masses are higher, which is expected since we removed the lowest mass objects from our fixed number density selection. At $z=0$ the evolving number density selection has 68 per cent fewer galaxies ($n(>M)=0.64\times10^{-3}$~Mpc~$^{-3}$). The evolving number density selection traces descendants much better. This is not true for lower number densities, including our fiducial value, as shown in Appendix~\ref{sec:evolve}. 

\begin{figure}
\center
\includegraphics[scale=.5]{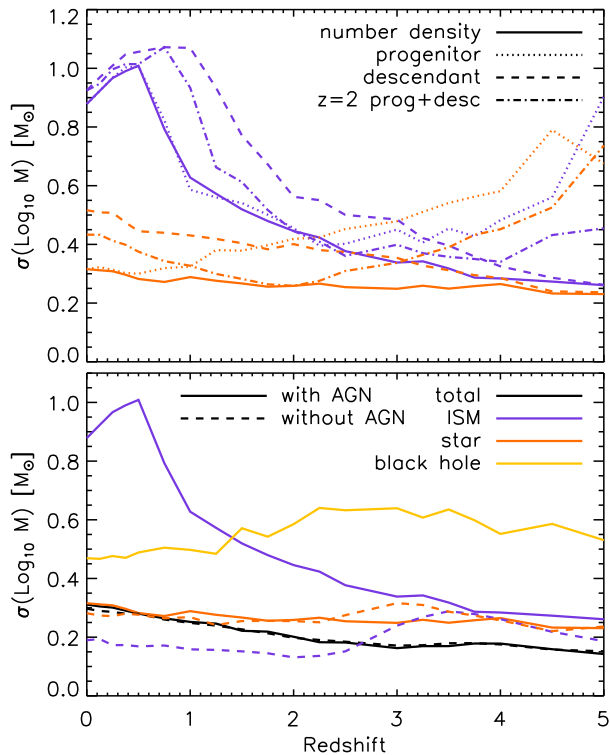}
\caption {\label{fig:scatter} \textit{Top panel:} Evolution of the scatter in the logarithm of the ISM mass (purple curves) and stellar mass (orange curves) for different methods to trace galaxies. Dotted (dashed) curves trace progenitors (descendants) from $z=0$ ($z=5$). Dot-dashed curves trace both progenitors and descendants from $z=2$. The scatter found for progenitors and descendants is similar to that found for fixed number densities, with the exception of increased scatter at high redshift for progenitors. \textit{Bottom panel:} Evolution of the scatter in the logarithm of the total halo mass (black curves), ISM mass (purple curves), stellar mass (orange curves), and black hole mass (yellow curves) at fixed number density with (solid curves) and without (dashed curves) AGN feedback (see Section~\ref{sec:compare} for discussion). The ISM mass scatter depends most on including AGN feedback. It increases strongly towards low redshift. The scatter in black hole mass is highest around $z=3$.}
\end{figure}

The top panel of Figure~\ref{fig:scatter} shows the 1$\sigma$ scatter, calculated by subtracting the logarithm of the 16th from the 84th percentile of the mass distribution and dividing by two, in the stellar masses (orange curves) and ISM masses (purple curves) for $n(>M)=2\times10^{-4}$~Mpc$^{-3}$. We compare the different galaxy selection methods: fixed number density (solid curves), progenitors of galaxies selected at $z=0$ (dotted curves), descendants of galaxies selected at $z=5$ (dashed curves), and progenitors and descendants of galaxies selected at $z=2$ (dot-dashed curves). Whereas the scatter in stellar masses increases only moderately ($\sim0.2-0.3$~dex) for fixed number density selection, the scatter for the descendant selection increases more strongly (by 0.2~dex) to 0.5~dex. The largest difference is found for the progenitor selection at high redshift, where it is as high as 0.7~dex, 0.5 dex above the one based on fixed number density. Tracing progenitors and descendants of $z=2$ galaxies also leads to increased scatter, which is generally lower by $\sim0.1$~dex than methods using just progenitors or descendants from $z=0-5$. 

At high redshift ($z>3$), the scatter in ISM mass is similar to the scatter in stellar mass for galaxies at fixed number density ($0.2-0.3$~dex). Again, the scatter for the progenitor selection deviates substantially towards higher redshift, reaching 0.9~dex at $z=5$. This indicates that a significant fraction of the progenitors of massive $z=0$ galaxies have different assembly histories than found through number density selections, at least at very high redshift. At low redshift ($z<3$) the scatter in ISM mass increases from 0.3 to 1~dex and exceeds the scatter in stellar mass by up to a factor of three for all selection methods\footnote{The scatter for the descendant tracing methods is generally larger, especially at $z\approx1$, yet it is very similar to the other methods at $z\lesssim0.5$ and follows the same trend with redshift.}. This result is therefore in large part insensitive to our method, but an inherent property of the galaxy population. As galaxies grow, a larger fraction of them become quenched, increasing the scatter in ISM mass. 

It seems surprising at first that a large scatter in ISM mass does not result in a large scatter in stellar mass. This can be understood by realizing that the ISM mass of a single galaxy is variable with time and will sometimes be above the median and sometimes below, depending on when a large galactic outflow happened, in the case of this simulation triggered by AGN activity. For individual galaxies the scatter averages out over time.

In the bottom panel of Figure~\ref{fig:scatter} we show total halo mass (black curves) and black hole mass (yellow curves) as well for the fixed number density selection. The difference between simulations with and without AGN feedback (solid and dashed curves) will be discussed in Section~\ref{sec:compare}. The scatter in halo mass increases from 0.15~dex at $z=5$ to 0.3~dex at $z=0$ for a fixed number density selection and reaches maximally 0.5~dex for the other selections. Halo mass exhibits the lowest scatter, because it is what we select on, but it is lower than the scatter in stellar mass by less than 0.1~dex at all redshifts. It is interesting to note that the scatter in black hole mass increases slightly after $z=5$, reaches a maximum around redshift~3, where AGN activity is highest, and subsequently decreases down to $z=0$, as opposed to the scatter in other masses. The peak scatter shifts to higher (lower) redshift for lower (higher) number densities, since AGN feedback becomes important when $M_\mathrm{halo}\approx10^{12}$~M$_\odot$. At high redshift, the scatter in black hole mass is large compared to the scatter in other masses, which is true for all number densities and selection methods.

\subsection{Evolution at fixed number density}

\begin{figure}
\center
\includegraphics[scale=.5]{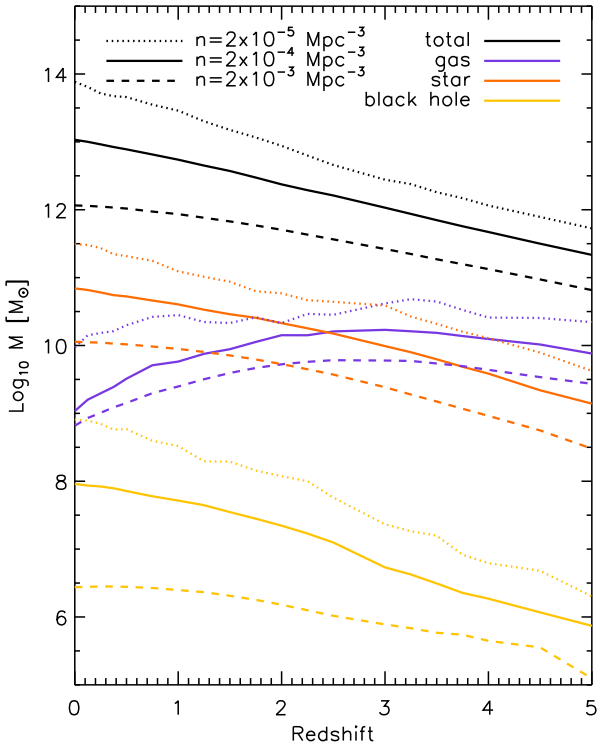}
\caption {\label{fig:nrs} Evolution of the median dark matter mass (black curves), ISM mass (purple curves), stellar mass (orange curves) and black hole mass (yellow curves) at different comoving number densities. Dotted, solid, and dashed curves show $n(>M)=2\times10^{-5}$~Mpc$^{-3}$, $2\times10^{-4}$~Mpc$^{-3}$, and $2\times10^{-4}$~Mpc$^{-3}$, respectively. The difference between different number densities are generally larger at lower redshift. The baryonic mass of the galaxy is dominated by the ISM at high redshift and by stars at low redshift. The redshift at which stars start to dominate over the ISM is lower for lower-mass galaxies.}
\end{figure}

Now that we have established that the evolution at fixed number density is similar (within a factor of a few) to the evolution of progenitors or descendants, we focus for the remainder of this paper on the evolution at fixed number density. This selection is most easily comparable to observations, but the results are similar for progenitors and descendants. Figure \ref{fig:nrs} shows the median evolution of the total halo mass (black curves), ISM mass (purple curves), stellar mass (orange curves), and black hole mass (yellow curves) from $z=0$ to $z=5$ at fixed number density increasing from $2\times10^{-5}$ (dotted curves) to $2\times10^{-4}$ (solid curves) to $2\times10^{-3}$~Mpc$^{-3}$ (dashed curves). For number densities different by a factor of 10, the halo masses differ by almost an order of magnitude at $z=0$ and by $\sim0.5$~dex at $z=5$. The difference between median stellar masses is about $\sim0.7$~dex. The difference in median ISM mass for $n(>M)=2\times10^{-4}$ and $n(>M)=2\times10^{-3}$~Mpc$^{-3}$ is small, because AGN feedback reduces the ISM mass in the more massive haloes, but is not so effective in lower-mass haloes. This is consistent with galaxies in haloes of $10^{12}$~M$_\odot$ being most efficient at forming stars. For the highest-mass haloes at low redshift, AGN feedback becomes less efficient at removing ISM gas and quenching star formation, consistent with observed star formation rates in brightest cluster galaxies.

At early times, the galaxy's mass is dominated by gas, but at late times stars dominate. The transition between ISM mass-dominated and stellar mass-dominated epoch happens at earlier times for more massive galaxies. It changes from $z=2.9$ to $z=2$ for $n(>M)=2\times10^{-5}$ to $2\times10^{-3}$~Mpc$^{-3}$. This shows that the progenitors of clusters become starved of gas at an earlier epoch than the progenitors of Milky Way-sized galaxies, consistent with the observed downsizing trend where more massive galaxies assemble earlier than less massive ones \citep[e.g.][]{Bundy2006, Perez2008, Marchesini2009, Mortlock2011, Muzzin2013}.

Observations show that the colours and specific star formation rates (SSFR) of galaxies in the Universe are bimodal \citep[e.g.][]{Kauffmann2003, Bell2004, Whitaker2011, Wuyts2011}. Following \citet{Franx2008}, we split our own sample into star-forming and quiescent galaxies by using $\mathrm{SSFR}<0.3/t_\mathrm{H}$ for quiescent galaxies, where $t_\mathrm{H}$ is the age of the Universe at the time the galaxy population is studied.

\begin{figure}
\center
\includegraphics[scale=.5]{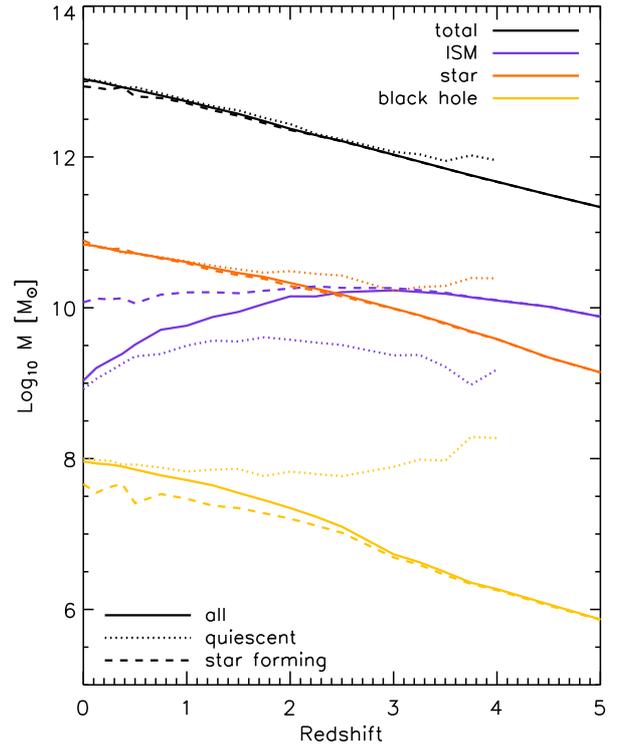}
\caption {\label{fig:qhist} Evolution of the median dark matter mass (black curves), ISM mass (purple curves), stellar mass (orange curves) and black hole mass (yellow curves) for all galaxies with $n(>M)=2\times10^{-4}$~Mpc$^3$ (solid curves) and for the subsample that is quiescent (dotted curves) or star-forming (dashed curves) at each redshift. The quiescent curves are truncated at $z=4$ because there are fewer than 10 quiescent galaxies at $z>4$. Quiescent galaxies have much lower ISM masses and much higher black hole masses than star-forming galaxies.}
\end{figure}

In Figure~\ref{fig:qhist} we show the median masses of galaxies subdivided into quiescent and star-forming galaxies at each redshift. The fraction of quiescent galaxies increases towards $z=0$ (as shown later quantitatively by the solid, black curve in Figure~\ref{fig:qf}). The quiescent curves are truncated at $z=4$ because there are fewer than 10 quiescent galaxies in our sample at higher redshift. Figure~\ref{fig:qhist} shows that quiescent galaxies have higher median total halo masses and stellar masses than star-forming galaxies, but this difference is small and disappears towards low redshift. Much bigger differences are found for ISM masses and black hole masses. The ISM mass of quiescent galaxies is about an order of magnitude lower than that of star-forming galaxies at $z=4$, which is partially due to the definition of quiescence being based on SSFR, which correlates with ISM mass. This difference decreases somewhat towards $z=1.5$ and increases again to 1~dex at $z=0$. Quiescent galaxies have, on average, black hole masses two orders of magnitude above the median black hole mass at $z=4$. At $z=0$ star-forming galaxies have black hole masses $\sim0.4$~dex below the median, in qualitative agreement with \citet{McConnell2013}. This shows that the mass of the central black hole anti-correlates with the ISM mass and thus with the star formation rate.

Quiescent galaxies do not necessarily stay quiescent in our simulations. Of the quiescent galaxies at $z=2$, 43 per cent were quiescent at $z=2.25$ and 69 per cent are still quiescent at $z=1.75$, so in our simulation there is considerable fluctuation in galaxies being quiescent and star-forming. This is a feature of the burstiness of AGN in our simulations.

\subsection{Accretion and star formation rates}

\begin{figure}
\center
\includegraphics[scale=.5]{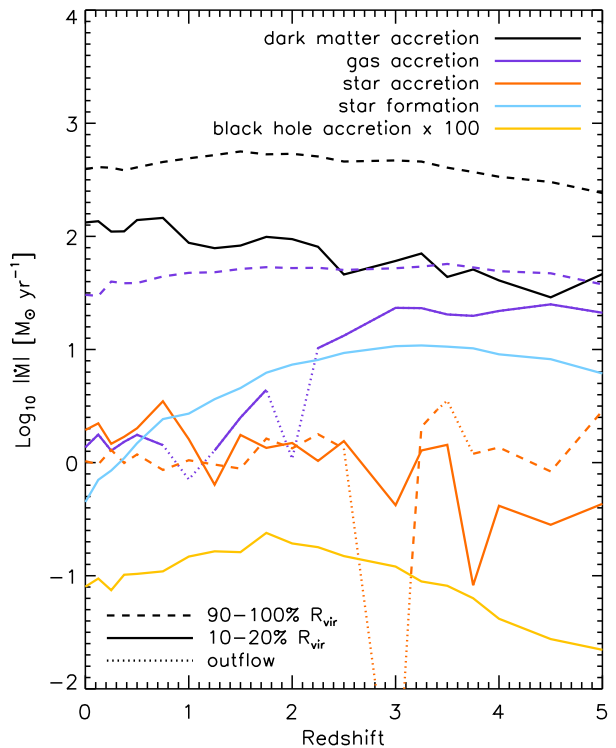}
\caption {\label{fig:flux} Dark matter (black), gas (purple), and star (orange) mass fluxes through a shell around the central galaxy at $0.9-1.0R_\mathrm{vir}$ (dashed curves) and $0.1-0.2R_\mathrm{vir}$ (solid curves). Outflows are shown as dotted segments. The star formation rate (blue) is measured inside $0.15R_\mathrm{vir}$. The black hole accretion rate (yellow) is multiplied by a factor 100 for visualization purposes. }
\end{figure}

In this subsection we focus on the time derivative of the mass. The appropriate definition of the inflow or outflow rate in an expanding Universe depends on the question of interest. Here we are interested in the mass growth of haloes in a comoving frame, where the haloes are defined using a criterion that would keep halo masses constant in time if there were no peculiar velocities, such as the spherical overdensity criterion we use here. The mass flux through a spherical shell with comoving radius $R$ is then calculated as follows:
\begin{equation}
\dot{M}_{\rm gas}(R) = \sum_{R \le r_i < R+dR} \dfrac{m_{i}v_{{\rm rad},i}}{V_\mathrm{shell}}A_\mathrm{shell},
\label{eq:mdotgas}
\end{equation}
Where $r_i$,  $m_i$, and $v_{{\rm rad},i}$ are the radius, mass, and radial velocity of particle $i$, and $dR$, $V_\mathrm{shell}$, and $A_\mathrm{shell}$ are the bin size, volume, and area (evaluated at $R+\frac{1}{2}dR$) of the shell. 

Figure \ref{fig:flux} shows median accretion rates and star formation rates for galaxies at a fixed number density $n(>M)=2\times10^{-4}$~Mpc$^{-3}$. To compute mass fluxes, we used the peculiar radial velocity of the gas in shells around the central galaxy, i.e.\ the Hubble flow is not taken into account. Dashed (solid) curves show the net inflow rate at $0.9-1R_\mathrm{vir}$ ($0.1-0.2R_\mathrm{vir}$). At fixed mass, accretion rates at the virial radius are much lower at lower redshift \citep[e.g.][]{Voort2011a, Faucher2011}, because the average density of the Universe, $\bar{\rho}$, decreases. However, at fixed number density, the accretion rates at $R_\mathrm{vir}$ do not decrease as strongly due to the increase in halo mass, which almost balances the decrease of $\bar{\rho}$. The number density shown in Figure \ref{fig:flux} is special in this regard. For $n(>M)=2\times10^{-5}$~Mpc$^{-3}$ ($2\times10^{-3}$~Mpc$^{-3}$) the dark matter, gas, and star accretion rates increase (decrease) somewhat towards $z=0$. When including the Hubble flow in our calculations of the velocity results, the accretion rate at $R_\mathrm{vir}$ at $z<1.5$ decreases and drops to zero by $z=0$. This is consistent with the picture of `pseudo-evolution' of dark matter haloes, i.e. that there is no growth in dark matter within a fixed physical radius at low redshift for $M_\mathrm{halo}\lesssim10^{13}$ \citep[e.g.][]{Diemer2013}.  

The net amount of gas flowing into the galaxy around $0.15R_\mathrm{vir}$ decreases strongly towards $z=0$ and is lower than the star formation rate at $z<2$, in agreement with observations \citep{Ownsworth2014}. This results in a decrease of the ISM mass. The gas accretion rate onto the galaxy is only a factor of two below the gas accretion rate onto the halo at $z>3$, but it is lower by $\sim1.4$~dex at $z<1$. At $z=1$ and $z=2$ we even find a net outflow rate, meaning that more than half of the galaxies in our sample are showing net outflow. This is due to AGN feedback resulting in large outflows that leave the central galaxy. These outflows do not necessarily leave the halo and we see no outflow in the median gas flux at $0.95R_\mathrm{vir}$. When including the Hubble flow to calculate gas velocities, the gas accretion rate at $R_\mathrm{vir}$ decreases at low redshift, but less steeply than that of the dark matter. At $0.15R_\mathrm{vir}$, on the other hand, we find average outflows at all $z\lesssim2$, decreasing from 5 to 0.5~M$_\odot$yr$^{-1}$. \citet{Wetzel2015} found physical accretion for gas at all radii (as opposed to dark matter), because gas is able to cool radiatively. The reason we instead find physical outflow is that our simulations include strong AGN feedback, which their simulations did not.

The star formation rate peaks around $z=3$, after which AGN feedback reduces it substantially. 
The amount of stars accreting onto the halo at $R_\mathrm{vir}$ is approximately the same as onto the galaxy at $0.15R_\mathrm{vir}$ at about 1~M$_\odot$yr$^{-1}$. These stars are brought in by mergers, of any mass ratio. Previous studies have found that most mass is brought in through minor mergers with mass ratios close to 1:10 \citep[e.g.][]{Purcell2007, Puchwein2010} and the same is expected to be true for these simulations.

For our fiducial number density, the growth rate due to mergers (i.e.\ the star accretion rate) is relatively constant at $z<3$, qualitatively consistent with \citet{Ownsworth2014}. It is important to notice that at $z\lesssim1$ the star accretion rate exceeds both the gas accretion rate and the star formation rate and becomes the dominant mode for the stellar mass growth of these galaxies as also seen in observations \citep[e.g.][]{Ownsworth2014, Vulcani2016} and other simulations \citep[e.g.][]{Oser2010}. For higher mass galaxies ($n(>M)=2\times10^{-5}$~Mpc$^{-3}$) the stellar mass growth is dominated by mergers at $z<2$, but for lower mass galaxies ($n(>M)=2\times10^{-3}$~Mpc$^{-3}$) star formation remains dominant down to $z=0$.
The behaviour of progenitors and descendants is qualitatively the same, with differences of only $0.2-0.4$~dex. 

The accretion rate onto the central black hole increases towards $z=1.75$ and then decreases by a factor of 3 towards $z=0$. This is driven by the decrease in central gas density, which is caused by feedback from the black hole itself.

We note that these accretion rates are median instantaneous accretion rates. Rare (high or low) accretion rates caused by major mergers or feedback, for example, will not affect the median accretion rate much, but they could affect the median mass growth, which represents a time averaged accretion rate. We therefore also calculated the average mass growth rate based on mass differences in Figure~\ref{fig:nrs}, indeed finding higher values for gas and dark accretion up to a factor of two, depending on redshift, likely caused by (rare) mergers. Much larger differences, up to an order of magnitude, are found for the dark matter accretion rate, which can be due to mergers as well. An alternative possibility is that strong AGN-driven outflows result in very bursty black hole accretion and high accretion rates are rare but contribute strongly to the growth of supermassive black holes. This is consistent with large scatter in (instantaneous) black hole accretion rates of up to 2~dex. The main mode of black hole growth may also vary with redshift.

\section{Effect of AGN feedback} \label{sec:compare}

Above we showed the evolution for a simulation from the OWLS project, which has been found to be the most realistic for massive galaxies \citep{Schaye2010, McCarthy2010, McCarthy2011, Voort2011b}. However, no simulation is able to perfectly capture all necessary physical processes involved in the formation of galaxies. Using fixed comoving number densities to trace galaxies across time is a useful way to directly compare observations and simulations and assess how well the evolution of galaxies is reproduced in a particular simulation. To show how the conclusions reached above depend on the feedback model implemented, we redid our analysis for a simulation that is identical except that it does not include AGN feedback. Fixed number densities provide a way to quantify the effect AGN feedback has on the average growth of galaxies in our simulations. In Section \ref{sec:obs} we will show how well both of these simulations compare to existing observations at fixed number density. 

These cosmological simulations try to include the relevant physical processes for galaxy formation, but need to employ simplified subgrid recipes to model stellar and black hole feedback. In recent years, huge progress has been made in understanding the effects of these subgrid models and in capturing important properties of observed galaxies and their haloes \citep[e.g.][]{Schaye2010, Dave2011, Vogelsberger2014, Genel2014, Schaye2015}. The growth of structure is followed from the early Universe until the present day and simulations are therefore a powerful tool to study galaxy formation dynamically. However, there are also limitations to how accurately simulations can produce realistic galaxies and it is therefore vital to directly compare them to observations. Doing this enables us to both validate or invalidate aspects of simulations as well as help interpret observations of galaxies.

\begin{figure}
\center
\includegraphics[scale=.5]{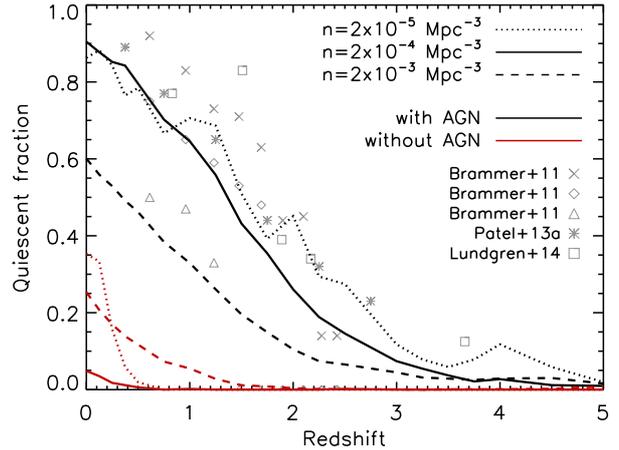}
\caption {\label{fig:qf} Fraction of quiescent galaxies as a function of redshift at fixed number density for a simulation with (black curves) and without (red curves). For all number densities, the fraction of quiescent galaxies is less than 30 per cent without AGN feedback. With AGN feedback 90 per cent of massive galaxies ($n(>M)<2\times10^{-4}$~Mpc$^{-3}$) at $z=0$ is quiescent. This is similar to the observed quiescent fractions \citep{Brammer2011, Patel2013a, Lundgren2014} as shown by the grey symbols (see text for a discussion).}
\end{figure}

Figure \ref{fig:qf} shows the quiescent fractions, defined as the fraction of galaxies with $\mathrm{SSFR}<0.3/t_\mathrm{H}$, as a function of redshift for a simulation with (black curves) and without (red curves) AGN feedback. Different linestyles show different number density selections, increasing from $2\times10^{-5}$~Mpc$^{-3}$ (dotted curves) to $2\times10^{-4}$~Mpc$^{-3}$ (solid curves) to $2\times10^{-3}$~Mpc$^{-3}$ (dashed curves). Including satellites and selecting on stellar mass reduces the quiescent fraction in the lowest number density bin, but has much less effect on our fiducial number density bin: at $z=0$, quiescent fractions are 0.63, 0.76, and 0.58 for $n(>M)=2\times10^{-5}$, $2\times10^{-4}$, and $2\times10^{-3}$~Mpc$^{-3}$, respectively, with AGN feedback. 

\begin{figure*}
\center
\includegraphics[scale=.59]{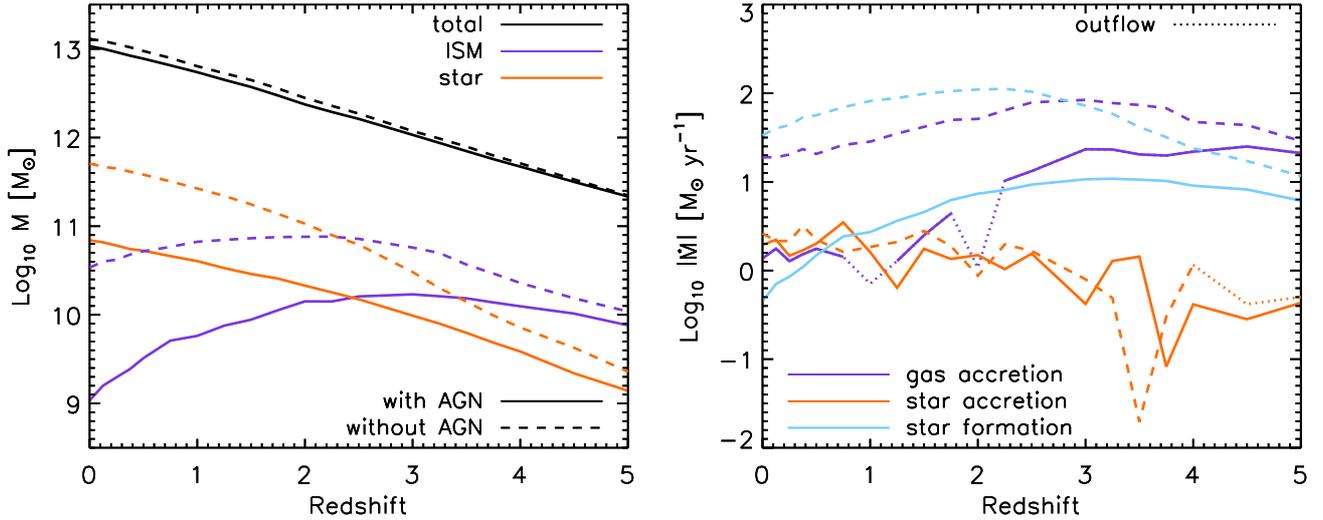}
\caption {\label{fig:REFAGN} Evolution of median masses and mass fluxes at fixed number density ($n(>M)=2\times10^{-4}$~Mpc$^{-3}$) for a simulation with (solid curves) and without (dashed curves) AGN feedback. Outflows are shown as dotted segments. Some curves are repeated from Figures~\ref{fig:nrs} (left) and \ref{fig:flux} (right). Colours are identical to the ones used in those Figures. AGN feedback decreases ISM and stellar masses, gas accretion rates, and star formation rates, most significantly when $M_\mathrm{halo}>10^{12}$~M$_\odot$, which happens at $z<3$ for this number density. The star accretion rates are not affected.}
\end{figure*}

It is immediately clear that in a simulation without AGN feedback, the vast majority of central galaxies are considered star-forming, even for the most massive objects. The fraction does increase to $\sim30$~per cent at $z=0$ for $n(>M)=2\times10^{-5}$ and $2\times10^{-3}$~Mpc$^{-3}$. This is likely due to the fact that these galaxies have turned most of their gas into stars and have thus depleted their own gas reservoir, which results in stellar masses higher than observed. With AGN feedback the two lowest number density selections have quiescent fraction increasing from $\lesssim0.1$ at $z=3$ to about 0.9 at $z=0$. For Milky-Way sized galaxies ($n(>M)=2\times10^{-3}$~Mpc$^{-3}$) the quiescent fraction increases to 60 per cent. The evolution of the quiescent fractions is in good agreement with \citet{Brammer2011}, \citet{Patel2013a}, and \citet{Lundgren2014}, with a steep increase after $z=2-3$ and low quiescent fractions at higher redshifts. The grey crosses, diamonds, and triangles show evolution at $M_\mathrm{star}=10^{11-11.6}$, $10^{10.6-11}$, and $10^{10.2-10.6}$~M$_\odot$, respectively. The grey asterisks (squares) show evolution in a 0.3~dex (0.2~dex) bin around $n=1.4\times10^{-4}$ ($n=3\times10^{-4}$). The simulated quiescent fractions are slightly lower than the observed ones, but, given the very different techniques used (for galaxy selection and analysis), the differences are minor. We conclude that the AGN simulation is more realistic when comparing to massive galaxies. 


The left panel of Figure \ref{fig:REFAGN} shows the total (black curves), gas (purple curves) and stellar (orange curves) mass evolution for central galaxies with $n(>M)=2\times10^{-4}$~Mpc$^{-3}$ for the simulation with (solid curves) and without (dashed curves) AGN feedback. The total halo mass of these haloes increases monotonically at approximately the same rate at all redshifts and does not depend on the inclusion of feedback. The difference is 0.1~dex by $z=0$. The ISM mass behaves very differently. It is strongly affected by AGN feedback. In both simulations, the ISM mass decreases at $z\lesssim2$, but this decrease is much stronger with AGN feedback implemented. Moreover, the accelerated growth that happens without AGN feedback at $z=3-4$, when the haloes are $\sim10^{12}$~M$_\odot$, is completely absent when AGN feedback is included. This has a strong effect on the evolution of the stellar mass. Whereas galaxy stellar masses increase steeply when their parent halo grows to $\sim10^{12}$~M$_\odot$ in a simulation without AGN feedback, galaxies grow at a slower rate when AGN feedback is included. The difference in stellar mass between the two simulations at $z=0$ is 0.9~dex, as opposed to 0.2~dex at $z=5$. The galaxies transition from being dominated by gas to being dominated by stars at the same redshift with or without AGN feedback. 

The difference in scatter in the mass evolution is shown in the bottom panel of Figure~\ref{fig:scatter} for simulations with (solid curves) and without (dashed curves) AGN feedback. The scatter in total halo mass and stellar mass is nearly identical (at about $0.2-0.3$~dex) in the two simulations. However, with AGN feedback the scatter in ISM mass increases from 0.3~dex at $z=5$ to 1~dex at $z=0.5$, significantly exceeding the scatter in stellar mass. Without AGN feedback it decreases around $z=3$ to less than 0.2~dex. 

The right panel of Figure \ref{fig:REFAGN} shows the gas accretion rate (purple curves) at 15 per cent of $R_\mathrm{vir}$ and the star formation rate (cyan curves) within the same radius. The inclusion of AGN feedback does not strongly impact the accretion of dark matter and stars, because both are collisionless and the stellar mass accreted is mostly from low-mass galaxies, which are not strongly affected by AGN feedback. The gas accretion rate, however, is reduced by an order of magnitude at $z=0$\footnote{The gas accretion rate at $R_\mathrm{vir}$ is reduced by maximally 0.3~dex by AGN feedback, so feedback affects the gas mostly inside the virial radius \citep{Voort2011a}.} and the star formation rate by almost two orders of magnitude. The magnitude of this effect is smaller at higher redshifts, but still appreciable over the entire redshift range probed here. With AGN feedback, the star accretion rate dominates at low redshift. This means that mergers are responsible for the majority of stellar mass growth. In simulations without AGN feedback, the gas accretion rate and star formation rate are an order of magnitude higher than the star accretion rate, meaning that the galaxy grows primarily through star formation.

\section{Comparison to observations} \label{sec:obs}

This work was partially motivated by the observations of \citet{Dokkum2010} and \citet{Papovich2011}. Both studies use a fixed number density, the former using mass bins at a fixed value of the stellar mass function $n=2\times10^{-4}$~Mpc$^{-3}$~dex$^{-1}$ and the latter all galaxies brighter than the luminosity, $L$, for which $n(>L)=2\times10^{-4}$~Mpc$^{-3}$. Both of these studies find higher stellar masses than our fiducial simulation. This will in part be due to the fact that the cosmological parameter of density fluctuations we used $\sigma_8=0.74$ is low compared to recent measurements that find $\sigma_8=0.82$ \citep{Hinshaw2013}. Additionally, our feedback is slightly too efficient at quenching star formation as it was not tuned to match the stellar mass function. Given these caveats and different ways of selecting galaxies, there is good agreement with the evolution found in our fiducial simulation, with stellar mass increasing by 0.5~dex since $z=2$. The simulation without AGN feedback, on the other hand, results in stellar masses that are too high and grow too fast. The ISM masses we find with AGN feedback are in reasonable agreement with those of \citet{Papovich2011}, whereas those without AGN feedback are too high (see Figure~\ref{fig:REFAGN}).

\citet{Muzzin2013} show about 0.5~dex growth since $z=2$ for galaxies at fixed number density with $M_\mathrm{star}(z=0)=10^{11}$~M$_\odot$, consistent with our results. However, they show faster stellar mass growth for lower-mass galaxies and slower mass growth for higher-mass galaxies (below $z=2$). Their most massive galaxies only grow by 0.2~dex since $z=2$, consistent with \citet{Brammer2011} and \citep{Marchesini2014}. This is in contrast with what we find at $n(>M)=2\times10^{-5}$~Mpc$^{-3}$. This could therefore be a manifestation of a problem with galaxy formation in clusters in our model. \citet{Leja2013} also find faster growth for more massive galaxies using a semi-analytic model, more consistent with our results. Simulations which match the observed stellar mass functions more closely, are able to better test this \citep{Vogelsberger2014, Schaye2015}. The evolutionary mass tracks found by \citet{Torrey2015} show an increase of about 0.5~dex since $z=2$ for massive galaxies at fixed number density, consistent with our results. In general, these authors find trends with redshift (including scatter in stellar mass) and differences between progenitors, descendants, and galaxies at fixed number density that are in quantitative agreement with ours. 

Comparing stellar mass growth at fixed and evolving number density, \citet{Papovich2015} find differences of up to 0.5~dex (for mass bins around $n=0.4-1.3\times10^{-3}$~Mpc$^{-3}$~dex$^{-1}$), similar to what we found for our highest number density selection (see Figure~\ref{fig:nrevol5141}). However, this is a larger difference than for our fiducial number density (see Figure~\ref{fig:nrevol514}).

As in \citet{Papovich2011}, we find that the ISM mass (gas accretion rate) exceeds the stellar mass (star formation rate) at high redshift. The transition to the stellar mass (star formation rate) dominated regime happens at somewhat earlier times in the observations. The trend with redshift is robust, which shows that the correct qualitative behaviour is captured in simulations. \citet{Ownsworth2014} show that star formation dominates the growth of galaxies at high redshift, whereas (minor) mergers dominate at low redshift. The gas accretion rate and star formation rate they derive decreases strongly below $z=2$, whereas the growth rate due to mergers remains relatively constant, consistent with our results (see Figure~\ref{fig:flux}).

Observational data points from the literature are shown in Appendix~\ref{sec:data} together with the stellar mass evolution in our fiducial simulation using different tracing methods at $n(>M)=2\times10^{-5}$, $2\times10^{-4}$, and $2\times10^{-3}$~Mpc$^{-3}$. 

\begin{figure}
\center
\includegraphics[scale=.5]{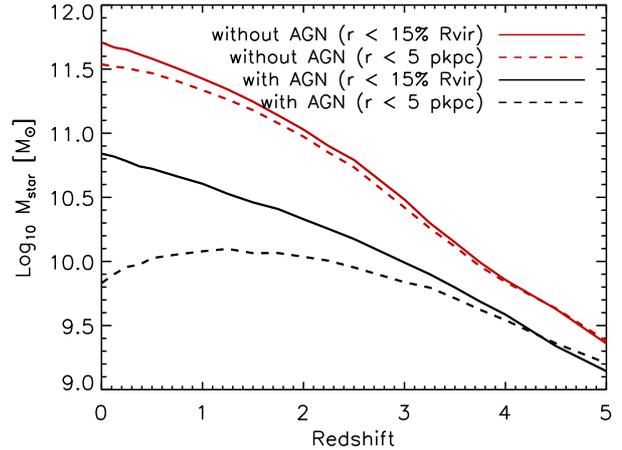}
\caption {\label{fig:5pkpc} Median stellar mass for $n(>M)=2\times10^{-4}$~Mpc$^{-3}$ within 0.15$R_\mathrm{vir}$ and 5~proper kpc. Black (red) curves show the evolution for galaxies with (without) AGN feedback. Simulations with AGN feedback show negligible growth in the central 5~kpc below $z=2$, although mass is still added to the outer parts of the galaxy. Without AGN feedback star formation in the central region drives most of the galaxy growth.}
\end{figure}
Minor, predominantly dry, mergers are thought to be responsible for growth of the most massive galaxies at late times, because these galaxies have low specific star formation rates, yet their sizes increase significantly towards low redshift \citep[e.g.][]{Guo2008, Franx2008}. The stellar mass added through mergers has been shown, using simulations, to contribute mostly to the outer part of the galaxies \citep[e.g.][]{Bournaud2007, Naab2009}. Observations also show evidence of this `inside-out' growth of galaxies \citep[e.g.][]{Franx2008, Bezanson2009}. \citet{Dokkum2010} and \citet{Patel2013a} showed that massive galaxies selected at fixed number density grow by 0.3-0.4~dex overall from $z\approx2$ to $z=0$, but the stellar mass within a fixed physical radius remains constant. We see the same behaviour in our simulations. Figure~\ref{fig:5pkpc} shows the total stellar mass growth of galaxies with $n(>M)=2\times10^{-4}$~Mpc$^{-3}$ within 0.15$R_\mathrm{vir}$ (solid curves) and within 5~proper~kpc (dashed curves). Black curves show the mass growth for our fiducial simulation and red curves show results for a simulation without AGN feedback. We see that, for the fiducial run, indeed no more mass is added to the inner region after $z=2$, but the total mass increases by 0.5~dex. This is consistent with little star formation in the centre and mass being added to the outer parts of the galaxy by mergers at low redshift \citep{Oser2010, Font2011}. The total stellar mass decreases slightly at low redshift. This is not due to stellar mass loss, which gives $\sim0.2$~dex offset at all redshifts. The same effect is present in the progenitor and descendant tracing methods. It could be a result of limited resolution, since the gravitational softening is 2.7~kpc below $z=2.9$. We checked that when increasing the radius to 10 kpc, the mass decrease is not present. 

Simulations without AGN feedback show very different behaviour. Because star formation rates remain high and peak in the centre, the stellar mass increases nearly as rapidly as the total galaxy mass within $0.15R_\mathrm{vir}$ down to $z=0$. This is consistent with the fact that the star formation rate is much higher than the rate at which stars accrete, as shown in Figure~\ref{fig:REFAGN}. 

For more massive galaxies ($n(>M)=2\times10^{-5}$~Mpc$^{-3}$) the central 5~proper~kpc region stops growing slightly earlier ($z=3$), because their star formation is quenched earlier. The general behaviour is very similar to that shown in Figure~\ref{fig:5pkpc}. For less massive galaxies ($n(>M)=2\times10^{-3}$~Mpc$^{-3}$) the difference between including or excluding AGN feedback is much smaller and the growth of the central region halts at a lower redshift ($z=1$).

\section{Summary and conclusions} \label{sec:concl}

While in simulations the growth of an individual galaxy can be followed throughout cosmic time, no such tracking is possible in observations. Statistically, one might expect the $N$ most massive galaxies at high redshift to evolve into the $N$ most massive galaxies at low redshift \citep[e.g.][]{Papovich2011} or that galaxies stay at the same (cumulative) number density as they grow \citep[e.g.][]{Dokkum2010, Patel2013a, Dokkum2013}. However, galaxy mergers and scatter in gas accretion rates, complicate this simplified picture \citep[e.g.][]{Leja2013, Behroozi2013b, Mundy2015}. Motivated by observational studies that investigate the redshift dependence of the properties of galaxies selected at a fixed, comoving cumulative number density, we trace progenitors and descendants in cosmological, hydrodynamical simulations from the OWLS project and compare the derived galaxy growth to the one derived using haloes at fixed number density, $n(>M)$. Below we summarize our results.

Our simulations are able to trace the progenitors of the majority of galaxies with $n(>M)\le2\times10^{-5}$~Mpc$^{-3}$ at $z=0$ out to $z=5$ or vice versa. Because of halo mergers between sample members, the number of descendants decreases dramatically towards low redshift, in the most extreme case by a factor of 3 at $z=0$ for galaxies with $n(>M)=2\times10^{-3}$~Mpc$^{-3}$ at $z=5$. The fraction of galaxies that are contained in both the number density and progenitor selection decreases towards higher redshift. About a quarter of the progenitors of the most massive $z=0$ galaxies are amongst the most massive galaxies at $z=5$. The progenitor and descendant samples are about 50 per cent complete over $\Delta z=2$. 

The median stellar, ISM, and halo masses can be different by as much as a factor of 3 when using fixed comoving number densities to select central galaxies versus tracing progenitors or descendants of these galaxies for the redshift range $z=0-5$. However, a similar difference exists between tracing progenitors or descendants due to mergers and scatter in accretion histories. Observationally there is no way to know which high-redshift galaxies will merge with larger galaxies and which galaxies will not. Mergers and scatter in the assembly of haloes conspire to get similar growth for descendants of galaxies selected at $z=5$ and number density selected galaxies. The same is not true for descendants of galaxies selected at $z=2$, which have higher masses than number density selected galaxies at $z=0$ due to the decreased star formation and thus increased importance of mergers at low redshift. Median black hole masses show differences of up to a factor of~5 between our various tracing methods. For high number densities ($n(>M)=2\times10^{-3}$~Mpc$^{-3}$), taking into account mergers by decreasing the number density traces descendants better than fixed number density, but this is not the case for $n(>M)\le2\times10^{-4}$~Mpc$^{-3}$.

The scatter in ISM mass is much larger than the scatter in stellar mass for massive galaxies when AGN feedback is important, e.g.\ at $z<3$ for $n(>M)=2\times10^{-4}$~Mpc~$^{-3}$. Interestingly, without AGN feedback, the scatter in ISM mass is even lower than in the scatter in stellar mass.
When tracing progenitors and descendants, the scatter is generally somewhat higher as compared to tracing galaxies at fixed number density. This is especially true for progenitors at high redshift, for which the scatter is higher by up to a factor of four. The scatter in black hole mass is always large and peaks around $z=3$ for our fiducial number density. 

The lower the number density, i.e.\ the more massive the average galaxy, the faster the median halo mass increases doubles. The same is true for the central black hole mass. This effect is less pronounced for ISM and stellar masses, because AGN feedback reduces gas accretion and star formation preferentially in high-mass objects. The mass evolution of quiescent and star-forming galaxies separately shows minor differences at $z<3$ in total halo and stellar masses, but the median ISM mass (black hole mass) of quiescent galaxies is substantially lower (higher) than that of star-forming galaxies at all redshifts.

The gas accretion rate at the virial radius follows the dark matter accretion rate closely, but is about an order of magnitude lower. This is not true close to the central galaxy. At high redshift, the average gas accretion rate is only a factor two lower at 0.15$R_\mathrm{vir}$ than at 0.95$R_\mathrm{vir}$. At $z=0$, however, it is lower by 1.3~dex. This large decrease is in part caused by longer gas cooling times in higher-mass, hotter haloes at a lower average density of the Universe, but primarily by AGN feedback. At certain intermediate redshifts ($z\sim1-2$), the median gas accretion rate is negative, meaning that for more than half our galaxy sample outflow dominates over inflow at 0.15$R_\mathrm{vir}$. We identify three regimes of galaxy growth. Gas accretion dominates until star formation takes over at $z=2$. Below $z=1$, both the gas accretion rate and the star formation rate have decreased sufficiently for stellar mass brought in by (minor) mergers to dominate the galaxy's mass growth. 

The gas accretion rate at $n(>M)=2\times10^{-4}$~Mpc$^{-3}$ peaks at high redshift ($z\ge5$) and then drops by more than an order of magnitude. The star formation rate peaks somewhat later ($z=3$) and follows the drop of the gas accretion rate. The black hole accretion rate peaks at an even lower redshift ($z=1.8$) and declines mildly by about 0.5~dex to $z=0$. The star accretion rate, i.e.\ the rate at which stars are brought in by mergers, shows no decline. Its evolution resembles that of the dark matter accretion rate, but lower by two orders of magnitude.

The fraction of quiescent galaxies increases down to $z=0$, consistent with observations \citep[e.g.][]{Brammer2011,Patel2013a,Lundgren2014}. However, without AGN feedback, the vast majority of galaxies remains star forming even at low redshift. In this case, the galaxies in our simulations experience accelerated growth in ISM mass, but especially in stellar mass when $M_\mathrm{halo}\approx10^{12}$~M$_\odot$. As a result, in the absence of AGN feedback, their stellar masses are almost an order of magnitude higher at $z=0$ for $n(>M)=2\times10^{-4}$~Mpc$^{-3}$. The galaxies still transition from being gas accretion dominated to being star formation dominated at $z=3$, but mergers never become important. Without AGN feedback, there is significant star formation in the centre at low redshift and this is where most of the stellar mass is added. When AGN feedback is included, massive galaxies grow only in their outskirts (inside-out growth), consistent with observations. 

Following galaxies consistently across a large redshift range is one of the main observational challenges. Within the uncertainties presented here, cumulative number densities are a simple tool with which we can follow galaxy growth. It traces the mass evolution as well as either true progenitors or descendants, since the median masses found with the latter two methods are also different by a factor of $\sim3$ (for $z=0-5$). Additionally, it provides an excellent way to select similar populations of galaxies in simulations and observations and to directly compare them.

\section*{Acknowledgements}

I am grateful to the referee for thorough and constructive comments and to Shannon Patel, Joop Schaye, Marcel van Daalen, Eric Bell, Marijn Franx, and Ann-Marie Madigan for useful discussions. It is my pleasure to thank Marcel van Daalen for providing the power spectrum necessary to make the finite volume correction. Thanks also to Joanne Cohn, Tim Davis, Joop Schaye, and Marcel van Daalen for comments on an earlier version of this manuscript. I would also like to mention the hospitality of IATE in C\'ordoba, Argentina, where this work was started. I wish to thank the OWLS team for the use of the simulations. The simulations presented here were run on Stella, the LOFAR BlueGene/L system in Groningen, on the Cosmology Machine at the Institute for Computational Cosmology in Durham as part of the Virgo Consortium research programme, and on Darwin in Cambridge. This work received financial support from the Marie Curie Initial Training Network CosmoComp (PITN-GA-2009-238356).

\bibliographystyle{mnras}
\bibliography{nrdens}

\begin{thebibliography}{}
\makeatletter
\relax
\def\mn@urlcharsother{\let\do\@makeother \do\$\do\&\do\#\do\^\do\_\do\%\do\~}
\def\mn@doi{\begingroup\mn@urlcharsother \@ifnextchar [ {\mn@doi@}
  {\mn@doi@[]}}
\def\mn@doi@[#1]#2{\def\@tempa{#1}\ifx\@tempa\@empty \href
  {http://dx.doi.org/#2} {doi:#2}\else \href {http://dx.doi.org/#2} {#1}\fi
  \endgroup}
\def\mn@eprint#1#2{\mn@eprint@#1:#2::\@nil}
\def\mn@eprint@arXiv#1{\href {http://arxiv.org/abs/#1} {{\tt arXiv:#1}}}
\def\mn@eprint@dblp#1{\href {http://dblp.uni-trier.de/rec/bibtex/#1.xml}
  {dblp:#1}}
\def\mn@eprint@#1:#2:#3:#4\@nil{\def\@tempa {#1}\def\@tempb {#2}\def\@tempc
  {#3}\ifx \@tempc \@empty \let \@tempc \@tempb \let \@tempb \@tempa \fi \ifx
  \@tempb \@empty \def\@tempb {arXiv}\fi \@ifundefined
  {mn@eprint@\@tempb}{\@tempb:\@tempc}{\expandafter \expandafter \csname
  mn@eprint@\@tempb\endcsname \expandafter{\@tempc}}}

\bibitem[\protect\citeauthoryear{{Bagla} \& {Prasad}}{{Bagla} \&
  {Prasad}}{2006}]{Bagla2006}
{Bagla} J.~S.,  {Prasad} J.,  2006, MNRAS, \href
  {http://adsabs.harvard.edu/abs/2006MNRAS.370..993B} {370, 993}

\bibitem[\protect\citeauthoryear{{Behroozi}, {Wechsler}  \&
  {Conroy}}{{Behroozi} et~al.}{2013a}]{Behroozi2013a}
{Behroozi} P.~S.,  {Wechsler} R.~H.,   {Conroy} C.,  2013a, ApJ, \href
  {http://adsabs.harvard.edu/abs/2013ApJ...770...57B} {770, 57}

\bibitem[\protect\citeauthoryear{{Behroozi}, {Marchesini}, {Wechsler},
  {Muzzin}, {Papovich}  \& {Stefanon}}{{Behroozi}
  et~al.}{2013b}]{Behroozi2013b}
{Behroozi} P.~S.,  {Marchesini} D.,  {Wechsler} R.~H.,  {Muzzin} A.,
  {Papovich} C.,   {Stefanon} M.,  2013b, \mn@doi [ApJL]
  {10.1088/2041-8205/777/1/L10}, \href
  {http://adsabs.harvard.edu/abs/2013ApJ...777L..10B} {777, L10}

\bibitem[\protect\citeauthoryear{{Bell} et~al.,}{{Bell}
  et~al.}{2004}]{Bell2004}
{Bell} E.~F.,  et~al., 2004, ApJ, \href
  {http://adsabs.harvard.edu/abs/2004ApJ...608..752B} {608, 752}

\bibitem[\protect\citeauthoryear{{Benson}, {Bower}, {Frenk}, {Lacey}, {Baugh}
  \& {Cole}}{{Benson} et~al.}{2003}]{Benson2003}
{Benson} A.~J.,  {Bower} R.~G.,  {Frenk} C.~S.,  {Lacey} C.~G.,  {Baugh} C.~M.,
    {Cole} S.,  2003, ApJ, \href
  {http://adsabs.harvard.edu/abs/2003ApJ...599...38B} {599, 38}

\bibitem[\protect\citeauthoryear{{Bezanson}, {van Dokkum}, {Tal}, {Marchesini},
  {Kriek}, {Franx}  \& {Coppi}}{{Bezanson} et~al.}{2009}]{Bezanson2009}
{Bezanson} R.,  {van Dokkum} P.~G.,  {Tal} T.,  {Marchesini} D.,  {Kriek} M.,
  {Franx} M.,   {Coppi} P.,  2009, ApJ, \href
  {http://adsabs.harvard.edu/abs/2009ApJ...697.1290B} {697, 1290}

\bibitem[\protect\citeauthoryear{{Bondi} \& {Hoyle}}{{Bondi} \&
  {Hoyle}}{1944}]{Bondi1944}
{Bondi} H.,  {Hoyle} F.,  1944, MNRAS, \href
  {http://adsabs.harvard.edu/abs/1944MNRAS.104..273B} {104, 273}

\bibitem[\protect\citeauthoryear{{Booth} \& {Schaye}}{{Booth} \&
  {Schaye}}{2009}]{Booth2009}
{Booth} C.~M.,  {Schaye} J.,  2009, MNRAS, \href
  {http://adsabs.harvard.edu/abs/2009MNRAS.398...53B} {398, 53}

\bibitem[\protect\citeauthoryear{{Booth} \& {Schaye}}{{Booth} \&
  {Schaye}}{2011}]{Booth2011}
{Booth} C.~M.,  {Schaye} J.,  2011, MNRAS, \href
  {http://adsabs.harvard.edu/abs/2011MNRAS.413.1158B} {413, 1158}

\bibitem[\protect\citeauthoryear{{Bournaud}, {Jog}  \& {Combes}}{{Bournaud}
  et~al.}{2007}]{Bournaud2007}
{Bournaud} F.,  {Jog} C.~J.,   {Combes} F.,  2007, A\&A, \href
  {http://adsabs.harvard.edu/abs/2007A%26A...476.1179B} {476, 1179}

\bibitem[\protect\citeauthoryear{{Brammer}, {Whitaker}, {van Dokkum}  \& {et
  al.}}{{Brammer} et~al.}{2011}]{Brammer2011}
{Brammer} G.~B.,  {Whitaker} K.~E.,  {van Dokkum} P.~G.,   {et al.} 2011, ApJ,
  \href {http://adsabs.harvard.edu/abs/2011ApJ...739...24B} {739, 24}

\bibitem[\protect\citeauthoryear{{Bryan} \& {Norman}}{{Bryan} \&
  {Norman}}{1998}]{Bryan1998}
{Bryan} G.~L.,  {Norman} M.~L.,  1998, ApJ, \href
  {http://adsabs.harvard.edu/abs/1998ApJ...495...80B} {495, 80}

\bibitem[\protect\citeauthoryear{{Bundy}, {Ellis}, {Conselice}  \& {et
  al.}}{{Bundy} et~al.}{2006}]{Bundy2006}
{Bundy} K.,  {Ellis} R.~S.,  {Conselice} C.~J.,   {et al.} 2006, ApJ, \href
  {http://adsabs.harvard.edu/abs/2006ApJ...651..120B} {651, 120}

\bibitem[\protect\citeauthoryear{{Chabrier}}{{Chabrier}}{2003}]{Chabrier2003}
{Chabrier} G.,  2003, PASP, \href
  {http://adsabs.harvard.edu/abs/2003PASP..115..763C} {115, 763}

\bibitem[\protect\citeauthoryear{{Clauwens}, {Franx}  \& {Schaye}}{{Clauwens}
  et~al.}{2016}]{Clauwens2016}
{Clauwens} B.,  {Franx} M.,   {Schaye} J.,  2016, preprint, \href
  {http://adsabs.harvard.edu/abs/2016arXiv160500009C} {1605.00009}

\bibitem[\protect\citeauthoryear{{Conroy} \& {Wechsler}}{{Conroy} \&
  {Wechsler}}{2009}]{Conroy2009}
{Conroy} C.,  {Wechsler} R.~H.,  2009, ApJ, \href
  {http://adsabs.harvard.edu/abs/2009ApJ...696..620C} {696, 620}

\bibitem[\protect\citeauthoryear{{Crain}, {McCarthy}, {Frenk}, {Theuns}  \&
  {Schaye}}{{Crain} et~al.}{2010}]{Crain2010a}
{Crain} R.~A.,  {McCarthy} I.~G.,  {Frenk} C.~S.,  {Theuns} T.,   {Schaye} J.,
  2010, MNRAS, \href {http://adsabs.harvard.edu/abs/2010MNRAS.407.1403C} {407,
  1403}

\bibitem[\protect\citeauthoryear{{Crain}, {Schaye}, {Bower}  \& {et
  al.}}{{Crain} et~al.}{2015}]{Crain2015}
{Crain} R.~A.,  {Schaye} J.,  {Bower} R.~G.,   {et al.} 2015, \mn@doi [MNRAS]
  {10.1093/mnras/stv725}, \href
  {http://adsabs.harvard.edu/abs/2015MNRAS.450.1937C} {450, 1937}

\bibitem[\protect\citeauthoryear{{Dalla Vecchia} \& {Schaye}}{{Dalla Vecchia}
  \& {Schaye}}{2008}]{Vecchia2008}
{Dalla Vecchia} C.,  {Schaye} J.,  2008, MNRAS, \href
  {http://adsabs.harvard.edu/abs/2008MNRAS.387.1431D} {387, 1431}

\bibitem[\protect\citeauthoryear{{Damen}, {Labb{\'e}}, {Franx}, {van Dokkum},
  {Taylor}  \& {Gawiser}}{{Damen} et~al.}{2009}]{Damen2009}
{Damen} M.,  {Labb{\'e}} I.,  {Franx} M.,  {van Dokkum} P.~G.,  {Taylor} E.~N.,
    {Gawiser} E.~J.,  2009, ApJ, \href
  {http://adsabs.harvard.edu/abs/2009ApJ...690..937D} {690, 937}

\bibitem[\protect\citeauthoryear{{Dav{\'e}}, {Oppenheimer}  \&
  {Finlator}}{{Dav{\'e}} et~al.}{2011}]{Dave2011}
{Dav{\'e}} R.,  {Oppenheimer} B.~D.,   {Finlator} K.,  2011, \mn@doi [MNRAS]
  {10.1111/j.1365-2966.2011.18680.x}, \href
  {http://adsabs.harvard.edu/abs/2011MNRAS.415...11D} {415, 11}

\bibitem[\protect\citeauthoryear{{Diemer}, {More}  \& {Kravtsov}}{{Diemer}
  et~al.}{2013}]{Diemer2013}
{Diemer} B.,  {More} S.,   {Kravtsov} A.~V.,  2013, \mn@doi [ApJ]
  {10.1088/0004-637X/766/1/25}, \href
  {http://adsabs.harvard.edu/abs/2013ApJ...766...25D} {766, 25}

\bibitem[\protect\citeauthoryear{{Dolag}, {Borgani}, {Murante}  \&
  {Springel}}{{Dolag} et~al.}{2009}]{Dolag2009}
{Dolag} K.,  {Borgani} S.,  {Murante} G.,   {Springel} V.,  2009, MNRAS, \href
  {http://adsabs.harvard.edu/abs/2009MNRAS.399..497D} {399, 497}

\bibitem[\protect\citeauthoryear{{Faucher-Gigu{\`e}re}, {Kere{\v s}}  \&
  {Ma}}{{Faucher-Gigu{\`e}re} et~al.}{2011}]{Faucher2011}
{Faucher-Gigu{\`e}re} C.-A.,  {Kere{\v s}} D.,   {Ma} C.-P.,  2011, MNRAS,
  \href {http://adsabs.harvard.edu/abs/2011MNRAS.417.2982F} {417, 2982}

\bibitem[\protect\citeauthoryear{{Font}, {McCarthy}, {Crain}, {Theuns},
  {Schaye}, {Wiersma}  \& {Dalla Vecchia}}{{Font} et~al.}{2011}]{Font2011}
{Font} A.~S.,  {McCarthy} I.~G.,  {Crain} R.~A.,  {Theuns} T.,  {Schaye} J.,
  {Wiersma} R.~P.~C.,   {Dalla Vecchia} C.,  2011, MNRAS, \href
  {http://adsabs.harvard.edu/abs/2011MNRAS.416.2802F} {416, 2802}

\bibitem[\protect\citeauthoryear{{Franx}, {van Dokkum}, {Schreiber}, {Wuyts},
  {Labb{\'e}}  \& {Toft}}{{Franx} et~al.}{2008}]{Franx2008}
{Franx} M.,  {van Dokkum} P.~G.,  {Schreiber} N.~M.~F.,  {Wuyts} S.,
  {Labb{\'e}} I.,   {Toft} S.,  2008, ApJ, \href
  {http://adsabs.harvard.edu/abs/2008ApJ...688..770F} {688, 770}

\bibitem[\protect\citeauthoryear{{Genel}, {Vogelsberger}, {Springel}  \& {et
  al.}}{{Genel} et~al.}{2014}]{Genel2014}
{Genel} S.,  {Vogelsberger} M.,  {Springel} V.,   {et al.} 2014, \mn@doi
  [MNRAS] {10.1093/mnras/stu1654}, \href
  {http://adsabs.harvard.edu/abs/2014MNRAS.445..175G} {445, 175}

\bibitem[\protect\citeauthoryear{{Genzel}, {Newman}, {Jones}  \& {et
  al.}}{{Genzel} et~al.}{2011}]{Genzel2011}
{Genzel} R.,  {Newman} S.,  {Jones} T.,   {et al.} 2011, ApJ, \href
  {http://adsabs.harvard.edu/abs/2011ApJ...733..101G} {733, 101}

\bibitem[\protect\citeauthoryear{{Guo} \& {White}}{{Guo} \&
  {White}}{2008}]{Guo2008}
{Guo} Q.,  {White} S.~D.~M.,  2008, MNRAS, \href
  {http://adsabs.harvard.edu/abs/2008MNRAS.384....2G} {384, 2}

\bibitem[\protect\citeauthoryear{{Haardt} \& {Madau}}{{Haardt} \&
  {Madau}}{2001}]{Haardt2001}
{Haardt} F.,  {Madau} P.,  2001, in {Neumann} D.~M.,  {Tran} J.~T.~V.,  eds,
  Clusters of Galaxies and the High Redshift Universe Observed in X-rays.

\bibitem[\protect\citeauthoryear{{Haas}, {Schaye}, {Booth}, {Dalla Vecchia},
  {Springel}, {Theuns}  \& {Wiersma}}{{Haas} et~al.}{2013}]{Haas2013}
{Haas} M.~R.,  {Schaye} J.,  {Booth} C.~M.,  {Dalla Vecchia} C.,  {Springel}
  V.,  {Theuns} T.,   {Wiersma} R.~P.~C.,  2013, \mn@doi [MNRAS]
  {10.1093/mnras/stt1487}, \href
  {http://adsabs.harvard.edu/abs/2013MNRAS.435.2931H} {435, 2931}

\bibitem[\protect\citeauthoryear{{Hinshaw}, {Larson}, {Komatsu}  \& {et
  al.}}{{Hinshaw} et~al.}{2013}]{Hinshaw2013}
{Hinshaw} G.,  {Larson} D.,  {Komatsu} E.,   {et al.} 2013, ApJS, \href
  {http://adsabs.harvard.edu/abs/2013ApJS..208...19H} {208, 19}

\bibitem[\protect\citeauthoryear{{Kauffmann}, {Heckman}, {White}  \& {et
  al.,}}{{Kauffmann} et~al.}{2003}]{Kauffmann2003}
{Kauffmann} G.,  {Heckman} T.~M.,  {White} S.~D.~M.,   {et al.,} 2003, MNRAS,
  \href {http://adsabs.harvard.edu/abs/2003MNRAS.341...54K} {341, 54}

\bibitem[\protect\citeauthoryear{{Kennicutt}}{{Kennicutt}}{1998}]{Kennicutt1998}
{Kennicutt} Jr. R.~C.,  1998, ApJ, \href
  {http://adsabs.harvard.edu/abs/1998ApJ...498..541K} {498, 541}

\bibitem[\protect\citeauthoryear{{Leja}, {van Dokkum}  \& {Franx}}{{Leja}
  et~al.}{2013}]{Leja2013}
{Leja} J.,  {van Dokkum} P.,   {Franx} M.,  2013, \mn@doi [ApJ]
  {10.1088/0004-637X/766/1/33}, \href
  {http://adsabs.harvard.edu/abs/2013ApJ...766...33L} {766, 33}

\bibitem[\protect\citeauthoryear{{Lundgren}, {van Dokkum}, {Franx}  \& {et
  al.}}{{Lundgren} et~al.}{2014}]{Lundgren2014}
{Lundgren} B.~F.,  {van Dokkum} P.,  {Franx} M.,   {et al.} 2014, ApJ, \href
  {http://adsabs.harvard.edu/abs/2014ApJ...780...34L} {780, 34}

\bibitem[\protect\citeauthoryear{{Marchesini}, {van Dokkum}, {F{\"o}rster
  Schreiber}, {Franx}, {Labb{\'e}}  \& {Wuyts}}{{Marchesini}
  et~al.}{2009}]{Marchesini2009}
{Marchesini} D.,  {van Dokkum} P.~G.,  {F{\"o}rster Schreiber} N.~M.,  {Franx}
  M.,  {Labb{\'e}} I.,   {Wuyts} S.,  2009, \mn@doi [ApJ]
  {10.1088/0004-637X/701/2/1765}, \href
  {http://adsabs.harvard.edu/abs/2009ApJ...701.1765M} {701, 1765}

\bibitem[\protect\citeauthoryear{{Marchesini}, {Muzzin}, {Stefanon}  \& et
  al.}{{Marchesini} et~al.}{2014}]{Marchesini2014}
{Marchesini} D.,  {Muzzin} A.,  {Stefanon} M.,   et al. 2014, preprint, \href
  {http://adsabs.harvard.edu/abs/2014arXiv1402.0003M} {1402.0003}

\bibitem[\protect\citeauthoryear{{McCarthy}, {Schaye}, {Ponman}  \& {et
  al.,}}{{McCarthy} et~al.}{2010}]{McCarthy2010}
{McCarthy} I.~G.,  {Schaye} J.,  {Ponman} T.~J.,   {et al.,} 2010, MNRAS, \href
  {http://adsabs.harvard.edu/abs/2010MNRAS.406..822M} {406, 822}

\bibitem[\protect\citeauthoryear{{McCarthy}, {Schaye}, {Bower}
  et~al.}{{McCarthy} et~al.}{2011}]{McCarthy2011}
{McCarthy} I.~G.,  {Schaye} J.,  {Bower} R.~G.,   et~al., 2011, MNRAS, in
  press, arXiv:1008.4799, \href
  {http://adsabs.harvard.edu/abs/2011MNRAS.tmp...35M} {}

\bibitem[\protect\citeauthoryear{{McCarthy}, {Schaye}, {Font}, {Theuns},
  {Frenk}, {Crain}  \& {Dalla Vecchia}}{{McCarthy} et~al.}{2012}]{McCarthy2012}
{McCarthy} I.~G.,  {Schaye} J.,  {Font} A.~S.,  {Theuns} T.,  {Frenk} C.~S.,
  {Crain} R.~A.,   {Dalla Vecchia} C.,  2012, \mn@doi [MNRAS]
  {10.1111/j.1365-2966.2012.21951.x}, \href
  {http://adsabs.harvard.edu/abs/2012MNRAS.427..379M} {427, 379}

\bibitem[\protect\citeauthoryear{{McConnell} \& {Ma}}{{McConnell} \&
  {Ma}}{2013}]{McConnell2013}
{McConnell} N.~J.,  {Ma} C.-P.,  2013, \mn@doi [ApJ]
  {10.1088/0004-637X/764/2/184}, \href
  {http://adsabs.harvard.edu/abs/2013ApJ...764..184M} {764, 184}

\bibitem[\protect\citeauthoryear{{Mo}, {van den Bosch}  \& {White}}{{Mo}
  et~al.}{2010}]{Mo2010}
{Mo} H.,  {van den Bosch} F.~C.,   {White} S.,  2010, {Galaxy Formation and
  Evolution}.
Cambridge University Press

\bibitem[\protect\citeauthoryear{{Mortlock}, {Conselice}, {Bluck}, {Bauer},
  {Gr{\"u}tzbauch}, {Buitrago}  \& {Ownsworth}}{{Mortlock}
  et~al.}{2011}]{Mortlock2011}
{Mortlock} A.,  {Conselice} C.~J.,  {Bluck} A.~F.~L.,  {Bauer} A.~E.,
  {Gr{\"u}tzbauch} R.,  {Buitrago} F.,   {Ownsworth} J.,  2011, MNRAS, \href
  {http://adsabs.harvard.edu/abs/2011MNRAS.413.2845M} {413, 2845}

\bibitem[\protect\citeauthoryear{{Moster}, {Naab}  \& {White}}{{Moster}
  et~al.}{2013}]{Moster2013}
{Moster} B.~P.,  {Naab} T.,   {White} S.~D.~M.,  2013, MNRAS, \href
  {http://adsabs.harvard.edu/abs/2013MNRAS.428.3121M} {428, 3121}

\bibitem[\protect\citeauthoryear{{Mundy}, {Conselice}  \& {Ownsworth}}{{Mundy}
  et~al.}{2015}]{Mundy2015}
{Mundy} C.~J.,  {Conselice} C.~J.,   {Ownsworth} J.~R.,  2015, \mn@doi [MNRAS]
  {10.1093/mnras/stv860}, \href
  {http://adsabs.harvard.edu/abs/2015MNRAS.450.3696M} {450, 3696}

\bibitem[\protect\citeauthoryear{{Muzzin}, {Marchesini}, {Stefanon}  \& {et
  al.}}{{Muzzin} et~al.}{2013}]{Muzzin2013}
{Muzzin} A.,  {Marchesini} D.,  {Stefanon} M.,   {et al.} 2013, ApJ, \href
  {http://adsabs.harvard.edu/abs/2013ApJ...777...18M} {777, 18}

\bibitem[\protect\citeauthoryear{{Naab}, {Johansson}  \& {Ostriker}}{{Naab}
  et~al.}{2009}]{Naab2009}
{Naab} T.,  {Johansson} P.~H.,   {Ostriker} J.~P.,  2009, ApJL, \href
  {http://adsabs.harvard.edu/abs/2009ApJ...699L.178N} {699, L178}

\bibitem[\protect\citeauthoryear{{Oser}, {Ostriker}, {Naab}, {Johansson}  \&
  {Burkert}}{{Oser} et~al.}{2010}]{Oser2010}
{Oser} L.,  {Ostriker} J.~P.,  {Naab} T.,  {Johansson} P.~H.,   {Burkert} A.,
  2010, \mn@doi [ApJ] {10.1088/0004-637X/725/2/2312}, \href
  {http://adsabs.harvard.edu/abs/2010ApJ...725.2312O} {725, 2312}

\bibitem[\protect\citeauthoryear{{Ownsworth}, {Conselice}, {Mortlock},
  {Hartley}, {Almaini}, {Duncan}  \& {Mundy}}{{Ownsworth}
  et~al.}{2014}]{Ownsworth2014}
{Ownsworth} J.~R.,  {Conselice} C.~J.,  {Mortlock} A.,  {Hartley} W.~G.,
  {Almaini} O.,  {Duncan} K.,   {Mundy} C.~J.,  2014, \mn@doi [MNRAS]
  {10.1093/mnras/stu1802}, \href
  {http://adsabs.harvard.edu/abs/2014MNRAS.445.2198O} {445, 2198}

\bibitem[\protect\citeauthoryear{{Papovich}, {Finkelstein}, {Ferguson}, {Lotz}
  \& {Giavalisco}}{{Papovich} et~al.}{2011}]{Papovich2011}
{Papovich} C.,  {Finkelstein} S.~L.,  {Ferguson} H.~C.,  {Lotz} J.~M.,
  {Giavalisco} M.,  2011, MNRAS, \href
  {http://adsabs.harvard.edu/abs/2011MNRAS.412.1123P} {412, 1123}

\bibitem[\protect\citeauthoryear{{Papovich}, {Labb{\'e}}, {Quadri}  \& {et
  al.}}{{Papovich} et~al.}{2015}]{Papovich2015}
{Papovich} C.,  {Labb{\'e}} I.,  {Quadri} R.,   {et al.} 2015, \mn@doi [ApJ]
  {10.1088/0004-637X/803/1/26}, \href
  {http://adsabs.harvard.edu/abs/2015ApJ...803...26P} {803, 26}

\bibitem[\protect\citeauthoryear{{Patel}, {van Dokkum}, {Franx}  \& {et
  al.}}{{Patel} et~al.}{2013a}]{Patel2013a}
{Patel} S.~G.,  {van Dokkum} P.~G.,  {Franx} M.,   {et al.} 2013a, ApJ, \href
  {http://adsabs.harvard.edu/abs/2013ApJ...766...15P} {766, 15}

\bibitem[\protect\citeauthoryear{{Patel}, {Fumagalli}, {Franx}  \& {et
  al.}}{{Patel} et~al.}{2013b}]{Patel2013b}
{Patel} S.~G.,  {Fumagalli} M.,  {Franx} M.,   {et al.} 2013b, ApJ, \href
  {http://adsabs.harvard.edu/abs/2013ApJ...778..115P} {778, 115}

\bibitem[\protect\citeauthoryear{{P{\'e}rez-Gonz{\'a}lez}
  et~al.,}{{P{\'e}rez-Gonz{\'a}lez} et~al.}{2008}]{Perez2008}
{P{\'e}rez-Gonz{\'a}lez} P.~G.,  et~al., 2008, \mn@doi [ApJ] {10.1086/523690},
  \href {http://adsabs.harvard.edu/abs/2008ApJ...675..234P} {675, 234}

\bibitem[\protect\citeauthoryear{{Power} \& {Knebe}}{{Power} \&
  {Knebe}}{2006}]{Power2006}
{Power} C.,  {Knebe} A.,  2006, MNRAS, \href
  {http://adsabs.harvard.edu/abs/2006MNRAS.370..691P} {370, 691}

\bibitem[\protect\citeauthoryear{{Puchwein}, {Springel}, {Sijacki}  \&
  {Dolag}}{{Puchwein} et~al.}{2010}]{Puchwein2010}
{Puchwein} E.,  {Springel} V.,  {Sijacki} D.,   {Dolag} K.,  2010, MNRAS, \href
  {http://adsabs.harvard.edu/abs/2010MNRAS.406..936P} {406, 936}

\bibitem[\protect\citeauthoryear{{Purcell}, {Bullock}  \& {Zentner}}{{Purcell}
  et~al.}{2007}]{Purcell2007}
{Purcell} C.~W.,  {Bullock} J.~S.,   {Zentner} A.~R.,  2007, ApJ, \href
  {http://adsabs.harvard.edu/abs/2007ApJ...666...20P} {666, 20}

\bibitem[\protect\citeauthoryear{{Reed}, {Bower}, {Frenk}, {Jenkins}  \&
  {Theuns}}{{Reed} et~al.}{2007}]{Reed2007}
{Reed} D.~S.,  {Bower} R.,  {Frenk} C.~S.,  {Jenkins} A.,   {Theuns} T.,  2007,
  MNRAS, \href {http://adsabs.harvard.edu/abs/2007MNRAS.374....2R} {374, 2}

\bibitem[\protect\citeauthoryear{{Sales}, {Navarro}, {Schaye}, {Dalla Vecchia},
  {Springel}  \& {Booth}}{{Sales} et~al.}{2010}]{Sales2010}
{Sales} L.~V.,  {Navarro} J.~F.,  {Schaye} J.,  {Dalla Vecchia} C.,  {Springel}
  V.,   {Booth} C.~M.,  2010, MNRAS, \href
  {http://adsabs.harvard.edu/abs/2010MNRAS.409.1541S} {409, 1541}

\bibitem[\protect\citeauthoryear{{Schaye} \& {Dalla Vecchia}}{{Schaye} \&
  {Dalla Vecchia}}{2008}]{Schaye2008}
{Schaye} J.,  {Dalla Vecchia} C.,  2008, MNRAS, \href
  {http://adsabs.harvard.edu/abs/2007MNRAS.tmp.1159S} {383, 1210}

\bibitem[\protect\citeauthoryear{{Schaye}, {Dalla Vecchia}, {Booth}  \& {et
  al.,}}{{Schaye} et~al.}{2010}]{Schaye2010}
{Schaye} J.,  {Dalla Vecchia} C.,  {Booth} C.~M.,   {et al.,} 2010, MNRAS,
  \href {http://adsabs.harvard.edu/abs/2010MNRAS.402.1536S} {402, 1536}

\bibitem[\protect\citeauthoryear{{Schaye}, {Crain}, {Bower}  \& {et
  al.}}{{Schaye} et~al.}{2015}]{Schaye2015}
{Schaye} J.,  {Crain} R.~A.,  {Bower} R.~G.,   {et al.} 2015, \mn@doi [MNRAS]
  {10.1093/mnras/stu2058}, \href
  {http://adsabs.harvard.edu/abs/2015MNRAS.446..521S} {446, 521}

\bibitem[\protect\citeauthoryear{{Seljak} \& {Zaldarriaga}}{{Seljak} \&
  {Zaldarriaga}}{1996}]{Seljak1996}
{Seljak} U.,  {Zaldarriaga} M.,  1996, ApJ, \href
  {http://cdsads.u-strasbg.fr/abs/1996ApJ...469..437S} {469, 437}

\bibitem[\protect\citeauthoryear{{Sirko}}{{Sirko}}{2005}]{Sirko2005}
{Sirko} E.,  2005, ApJ, \href
  {http://adsabs.harvard.edu/abs/2005ApJ...634..728S} {634, 728}

\bibitem[\protect\citeauthoryear{{Springel}}{{Springel}}{2005}]{Springel2005}
{Springel} V.,  2005, MNRAS, \href
  {http://adsabs.harvard.edu/abs/2005MNRAS.364.1105S} {364, 1105}

\bibitem[\protect\citeauthoryear{{Springel} \& {Hernquist}}{{Springel} \&
  {Hernquist}}{2002}]{Springel2002}
{Springel} V.,  {Hernquist} L.,  2002, MNRAS, \href
  {http://cdsads.u-strasbg.fr/abs/2002MNRAS.333..649S} {333, 649}

\bibitem[\protect\citeauthoryear{{Springel}, {White}, {Tormen}  \&
  {Kauffmann}}{{Springel} et~al.}{2001}]{Springel2001}
{Springel} V.,  {White} S.~D.~M.,  {Tormen} G.,   {Kauffmann} G.,  2001, MNRAS,
  \href {http://adsabs.harvard.edu/abs/2001MNRAS.328..726S} {328, 726}

\bibitem[\protect\citeauthoryear{{Springel}, {Di Matteo}  \&
  {Hernquist}}{{Springel} et~al.}{2005}]{Springeletal2005}
{Springel} V.,  {Di Matteo} T.,   {Hernquist} L.,  2005, MNRAS, \href
  {http://adsabs.harvard.edu/abs/2005MNRAS.361..776S} {361, 776}

\bibitem[\protect\citeauthoryear{{Szomoru}, {Franx}  \& {van Dokkum}}{{Szomoru}
  et~al.}{2012}]{Szomoru2012}
{Szomoru} D.,  {Franx} M.,   {van Dokkum} P.~G.,  2012, ApJ, \href
  {http://adsabs.harvard.edu/abs/2012ApJ...749..121S} {749, 121}

\bibitem[\protect\citeauthoryear{{Torrey}, {Wellons}, {Machado}  \& {et
  al.}}{{Torrey} et~al.}{2015}]{Torrey2015}
{Torrey} P.,  {Wellons} S.,  {Machado} F.,   {et al.} 2015, \mn@doi [MNRAS]
  {10.1093/mnras/stv1986}, \href
  {http://adsabs.harvard.edu/abs/2015MNRAS.454.2770T} {454, 2770}

\bibitem[\protect\citeauthoryear{{van Daalen}, {Schaye}, {Booth}  \& {Dalla
  Vecchia}}{{van Daalen} et~al.}{2011}]{Daalen2011}
{van Daalen} M.~P.,  {Schaye} J.,  {Booth} C.~M.,   {Dalla Vecchia} C.,  2011,
  MNRAS, \href {http://adsabs.harvard.edu/abs/2011MNRAS.415.3649V} {415, 3649}

\bibitem[\protect\citeauthoryear{{van Dokkum}, {Whitaker}, {Brammer}  \& {et
  al.}}{{van Dokkum} et~al.}{2010}]{Dokkum2010}
{van Dokkum} P.~G.,  {Whitaker} K.~E.,  {Brammer} G.,   {et al.} 2010, ApJ,
  \href {http://adsabs.harvard.edu/abs/2010ApJ...709.1018V} {709, 1018}

\bibitem[\protect\citeauthoryear{{van Dokkum}, {Leja}, {Nelson}  \& et
  al.}{{van Dokkum} et~al.}{2013}]{Dokkum2013}
{van Dokkum} P.~G.,  {Leja} J.,  {Nelson} E.~J.,   et al. 2013, ApJL, \href
  {http://adsabs.harvard.edu/abs/2013ApJ...771L..35V} {771, L35}

\bibitem[\protect\citeauthoryear{{van de Voort}, {Schaye}, {Booth}, {Haas}  \&
  {Dalla Vecchia}}{{van de Voort} et~al.}{2011a}]{Voort2011a}
{van de Voort} F.,  {Schaye} J.,  {Booth} C.~M.,  {Haas} M.~R.,   {Dalla
  Vecchia} C.,  2011a, MNRAS, \href
  {http://adsabs.harvard.edu/abs/2011MNRAS.414.2458V} {414, 2458}

\bibitem[\protect\citeauthoryear{{van de Voort}, {Schaye}, {Booth}  \& {Dalla
  Vecchia}}{{van de Voort} et~al.}{2011b}]{Voort2011b}
{van de Voort} F.,  {Schaye} J.,  {Booth} C.~M.,   {Dalla Vecchia} C.,  2011b,
  MNRAS, \href {http://adsabs.harvard.edu/abs/2011MNRAS.415.2782V} {415, 2782}

\bibitem[\protect\citeauthoryear{{van de Voort}, {Schaye}, {Altay}  \&
  {Theuns}}{{van de Voort} et~al.}{2012}]{Voort2012}
{van de Voort} F.,  {Schaye} J.,  {Altay} G.,   {Theuns} T.,  2012, \mn@doi
  [MNRAS] {10.1111/j.1365-2966.2012.20487.x}, \href
  {http://adsabs.harvard.edu/abs/2012MNRAS.421.2809V} {421, 2809}

\bibitem[\protect\citeauthoryear{{Vogelsberger}, {Genel}, {Springel}  \& {et
  al.}}{{Vogelsberger} et~al.}{2014}]{Vogelsberger2014}
{Vogelsberger} M.,  {Genel} S.,  {Springel} V.,   {et al.} 2014, \mn@doi
  [MNRAS] {10.1093/mnras/stu1536}, \href
  {http://adsabs.harvard.edu/abs/2014MNRAS.444.1518V} {444, 1518}

\bibitem[\protect\citeauthoryear{{Vulcani}, {Marchesini}, {De Lucia}  \& {et
  al.}}{{Vulcani} et~al.}{2016}]{Vulcani2016}
{Vulcani} B.,  {Marchesini} D.,  {De Lucia} G.,   {et al.} 2016, \mn@doi [ApJ]
  {10.3847/0004-637X/816/2/86}, \href
  {http://adsabs.harvard.edu/abs/2016ApJ...816...86V} {816, 86}

\bibitem[\protect\citeauthoryear{{Wetzel} \& {Nagai}}{{Wetzel} \&
  {Nagai}}{2014}]{Wetzel2015}
{Wetzel} A.~R.,  {Nagai} D.,  2014, preprint, \href
  {http://adsabs.harvard.edu/abs/2014arXiv1412.0662W} {1412.0662}

\bibitem[\protect\citeauthoryear{{Whitaker}, {Labb{\'e}}, {van Dokkum}  \& {et
  al.}}{{Whitaker} et~al.}{2011}]{Whitaker2011}
{Whitaker} K.~E.,  {Labb{\'e}} I.,  {van Dokkum} P.~G.,   {et al.} 2011, ApJ,
  \href {http://adsabs.harvard.edu/abs/2011ApJ...735...86W} {735, 86}

\bibitem[\protect\citeauthoryear{{White}}{{White}}{1994}]{White1994}
{White} S.~D.~M.,  1994, preprint, \href
  {http://adsabs.harvard.edu/abs/1994astro.ph.10043W} {astro-ph/9410043}

\bibitem[\protect\citeauthoryear{{White} \& {Frenk}}{{White} \&
  {Frenk}}{1991}]{White1991}
{White} S.~D.~M.,  {Frenk} C.~S.,  1991, ApJ, \href
  {http://adsabs.harvard.edu/abs/1991ApJ...379...52W} {379, 52}

\bibitem[\protect\citeauthoryear{{Wiersma}, {Schaye}  \& {Smith}}{{Wiersma}
  et~al.}{2009a}]{Wiersma2009a}
{Wiersma} R.~P.~C.,  {Schaye} J.,   {Smith} B.~D.,  2009a, MNRAS, \href
  {http://adsabs.harvard.edu/abs/2009MNRAS.393...99W} {393, 99}

\bibitem[\protect\citeauthoryear{{Wiersma}, {Schaye}, {Theuns}, {Dalla Vecchia}
   \& {Tornatore}}{{Wiersma} et~al.}{2009b}]{Wiersma2009b}
{Wiersma} R.~P.~C.,  {Schaye} J.,  {Theuns} T.,  {Dalla Vecchia} C.,
  {Tornatore} L.,  2009b, MNRAS, \href
  {http://adsabs.harvard.edu/abs/2009MNRAS.399..574W} {399, 574}

\bibitem[\protect\citeauthoryear{{Wuyts}, {F{\"o}rster Schreiber}, {van der
  Wel}  \& {et al.}}{{Wuyts} et~al.}{2011}]{Wuyts2011}
{Wuyts} S.,  {F{\"o}rster Schreiber} N.~M.,  {van der Wel} A.,   {et al.} 2011,
  ApJ, \href {http://adsabs.harvard.edu/abs/2011ApJ...742...96W} {742, 96}

\bibitem[\protect\citeauthoryear{{Zel'Dovich}}{{Zel'Dovich}}{1970}]{Zeldovich1970}
{Zel'Dovich} Y.~B.,  1970, A\&A, \href
  {http://adsabs.harvard.edu/abs/1970A%26A.....5...84Z} {5, 84}


\makeatother
\end{thebibliography}

\bsp

\appendix

\section{Finite simulation volume correction} \label{sec:box}

In this work we are interested in studying the most massive galaxies and haloes in a simulation. Because of the finite volume of the simulation and its periodic boundary conditions, the halo mass function will not be the same as if the volume were infinite \citep[e.g.][]{Sirko2005, Bagla2006, Power2006, Reed2007}. To correct for this, we follow the method of \citet{Reed2007}. $\sigma(M)$ is the root-mean-square amplitude of the linearly-extrapolated density fluctuations smoothed with a top-hat filter with a radius that encloses a mass $M$ at the mean cosmic matter density. We calculate $\sigma^2(M)$ for the theoretical power spectrum and $\sigma^2(M')$ for the initial power spectrum measured from the simulations \citep{Daalen2011} and see for which masses $M$ and $M'$ they are equal. We correct the halo mass in the simulation to be the correct value $M$. Because of mass conservation, this means the number of haloes with mass $M$ is less than those with mass $M'$. The new halo is therefore weighted with the fraction $M'/M$ to calculate the number density and the median halo mass.

We repeated our analysis following the procedure described above to correct for the finite size of our simulation. Because of the fact that the corrected halo mass of these objects is higher, but their contribution to the total number density is lower, we include more haloes in our analysis, $N=75$, 687, and 6354 haloes, for number densities $n(>M)=2\times10^{-5}$, $2\times10^{-4}$, and $2\times10^{-3}$~Mpc$^{-3}$, respectively. The inclusion of lower mass objects balances out the higher masses of individual haloes and leads to very similar results. For simplicity, we therefore choose not to correct for the finite volume of our simulation for the rest of our results.

\section{Stellar mass, satellites, and mass bins} \label{sec:star}

\begin{figure}
\center
\includegraphics[scale=.5]{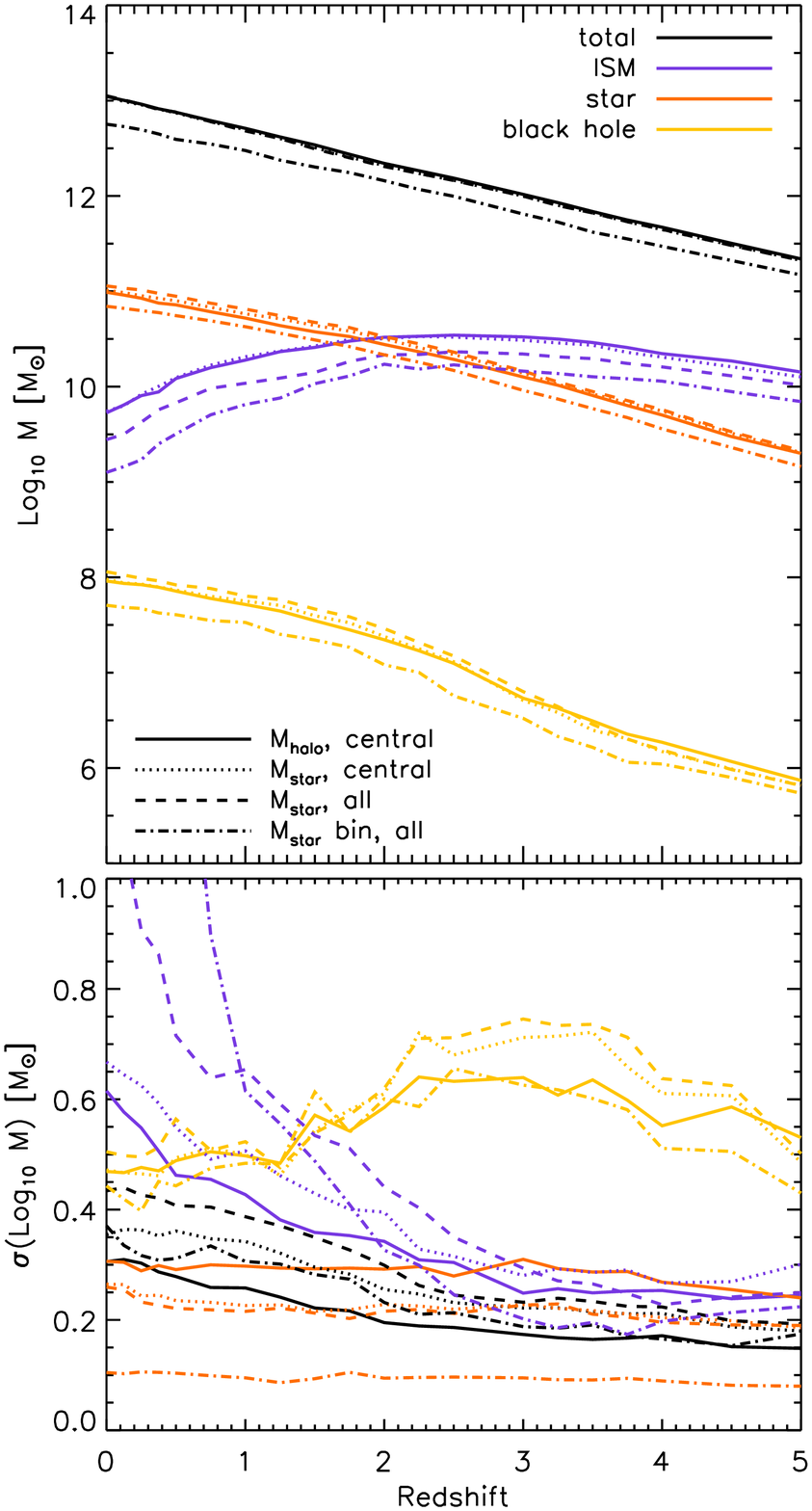}
\caption {\label{fig:histstar} Evolution of the median dark matter mass (black curves), ISM mass (purple curves), stellar mass (orange curves) and black hole mass (yellow curves) in the top panel and the evolution of the scatter in these masses in the bottom panel for $n(>M)=2\times10^{-4}$~Mpc$^{-3}$. The solid curves are the same as in Figure~\ref{fig:nrs} and Figure~\ref{fig:scatter}, respectively, ranking central galaxies based on their halo mass. Central galaxies are ranked according to their stellar mass for the dotted curves. We include all galaxies, centrals and satellites, and rank based on stellar mass for the dashed curves. The dot-dashed curves use the same ranking as the dashed curves, but only in a narrow mass bin around $n(>M)=2\times10^{-4}$~Mpc$^{-3}$. In general, the results are nearly identical, with the exception that using a mass bin reduces the masses by $0.2-0.4$~dex. The scatter in stellar mass decreases substantially when ranking on stellar mass and the scatter in ISM mass increases strongly at low redshift when including satellites.}
\end{figure}

We discussed our choice of ranking galaxies according to their halo mass in Section~\ref{sec:method} and showed its results in the subsequent sections. Figure \ref{fig:histstar} shows the difference we find when using different rankings. The top panel shows mass evolution as in Figures~\ref{fig:nrprogdesc} and~\ref{fig:nrs} and the bottom panel shows the evolution of mass scatter as in Figure~\ref{fig:scatter}. Solid curves use the same halo mass ranking, as before, but there is an important difference in determining stellar and ISM masses. Previously we included only the stellar mass and star-forming gas mass inside 15 per cent of $R_\mathrm{vir}$, but since the virial radius is not well-defined for satellites, here we include the stellar mass and star-forming gas mass in the entire main halo or subhalo for consistency.

Dotted curve shows the evolution if we rank galaxies according to their stellar mass instead of their halo mass, but still only include central galaxies. For the dashed curves we rank according to stellar mass, but include both centrals and satellites. In the dot-dashed curves central and satellite galaxies are ranked according to stellar mass, but only using the 100 galaxies with masses closest to that of galaxies with $n(>M)=2\times10^{-4}$~Mpc$^{-3}$, resulting in a narrower mass bin. For most rankings the results are very similar, with the exception that using a mass bin reduces the masses by $0.2-0.4$~dex, since we exclude the most massive galaxies and including more low-mass galaxies. The trends with redshift and mass are robust. \citet{Torrey2015} also find no significant difference between selecting number densities based on stellar or halo mass.

Because we study the evolution of massive galaxies, when we select on stellar mass and include all galaxies, we find that the sample is still dominated by centrals. On average, depending somewhat on redshift, centrals make up 94, 87, and 80 per cent of the population of galaxies with $n(>M)=2\times10^{-5}$, $2\times10^{-4}$, and $2\times10^{-3}$~Mpc$^{-3}$. Including satellites decreases the median ISM mass by up to 0.3~dex, but this is an artifact of the way we defined ISM masses in this plot, i.e.\ all star forming gas in a main halo or subhalo. Because the star-forming substructures that are too small to be identified by \textsc{subfind} are included in the main halo, its ISM mass will be artificially high.\footnote{The same is true for stellar masses, but the effect is much smaller, since most of the unidentified substructures do contain star-forming gas particles, but do not contain star particles.} We can see from Figure~\ref{fig:nrprogdesc} that the ISM mass of the central galaxies is much lower, by 0.3-0.7~dex, when including only the star-forming gas inside $0.15R_\mathrm{vir}$.

Unsurprisingly, the scatter in halo mass goes up and the scatter in stellar mass goes down when we rank galaxies based on their stellar masses. However, the difference is not very large, i.e.\ less than 0.1~dex. Including satellites increases the scatter in halo mass. The biggest difference is found when using only a mass bin around $n(>M)=2\times10^{-4}$~Mpc$^{-3}$, instead of including all massive galaxies, in which case the scatter in stellar mass decreases to 0.1~dex. The scatter in other masses remain high.

It is interesting to note that selecting on stellar mass, instead of halo mass, increases the scatter in black hole masses. However, the scatter in stellar mass decreases. This hints towards black hole growth being dominated by halo properties, rather than galaxy properties, even though the black hole accretion rates are set at much smaller scales \citep[see also][]{Booth2011}.

\section{Evolving comoving number density} \label{sec:evolve}

\begin{figure}
\center
\includegraphics[scale=.5]{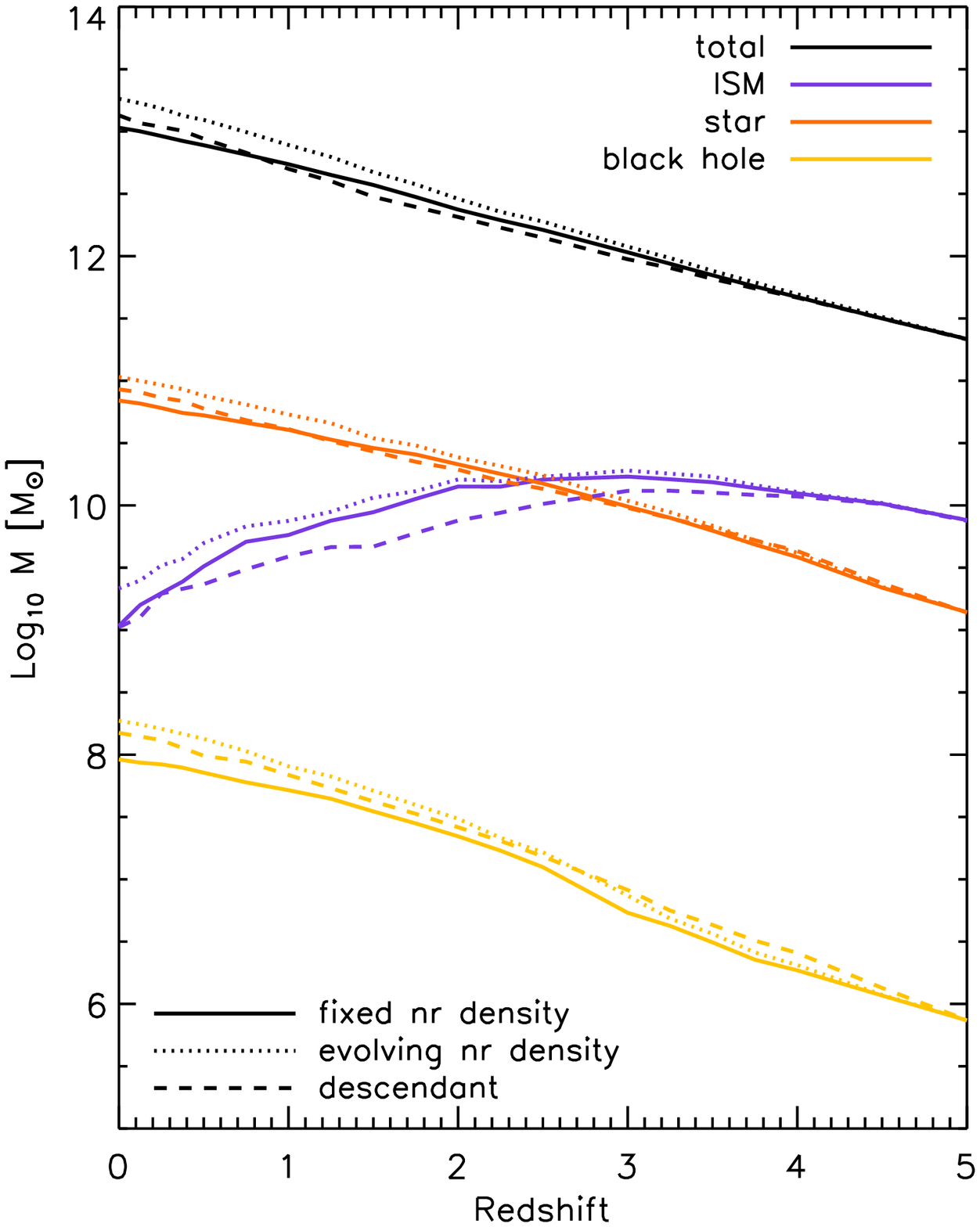}
\caption {\label{fig:nrevol514} Evolution of the median total halo mass (black curves), ISM mass (purple curves), stellar mass (orange curves) and black hole mass (yellow curves) using different methods to link high-redshift with low-redshift galaxies. The solid curves show median masses at fixed number density ($n(>M)=2\times10^{-4}$~Mpc$^{-3}$). Dotted curves use $z=5$ ($z=0$) galaxies with an evolving number density based on the number of mergers. Dashed curves curves use $z=5$ galaxies with the same number density, but trace their descendants down to $z=0$. The dotted and dashed curve therefore include the same number of galaxies. The mass evolution at fixed $n(>M)=2\times10^{-4}$~Mpc$^{-3}$ traces descendants more closely than the one at evolving number density.}
\end{figure}
Figure~\ref{fig:nrevol514} shows the same as Figure~\ref{fig:nrevol5141}, but for our fiducial number density ($n(>M)=2\times10^{-4}$~Mpc$^{-3}$), and includes the mass evolution at fixed number density (solid curves) and evolving number density (dotted) and for tracing $z=5$ descendants (dashed). Figure~\ref{fig:nrevol5141} showed that for $n(>M)=2\times10^{-3}$~Mpc$^{-3}$ using an evolving number density, which takes into account mergers, matches the descendant population much better than using a fixed number density. However, at $n(>M)=2\times10^{-4}$~Mpc$^{-3}$, the scatter in the growth history of galaxies (which decreases the median mass) and the number of mergers (which increases the median mass) approximately cancel out in such a way that a fixed number density selection traces the growth of galaxy descendants well. This is consistent with \citet{Torrey2015} who find no evolution in number density for galaxies with $M_\mathrm{star}(z=0)=10^{11}$~M$_\odot$ and a larger evolution for galaxies with higher number densities\footnote{Note that these authors use number density bins which result in lower masses for the same number density as shown in Figure\ref{fig:histstar}}. The difference between fixed and evolving number density is $0.2-0.3$ dex at $z=0$ and only 0.1~dex at $z=2$, so both are a reasonably accurate way to trace descendants.

\section{Observational data} \label{sec:data}

\begin{figure}
\center
\includegraphics[scale=.5]{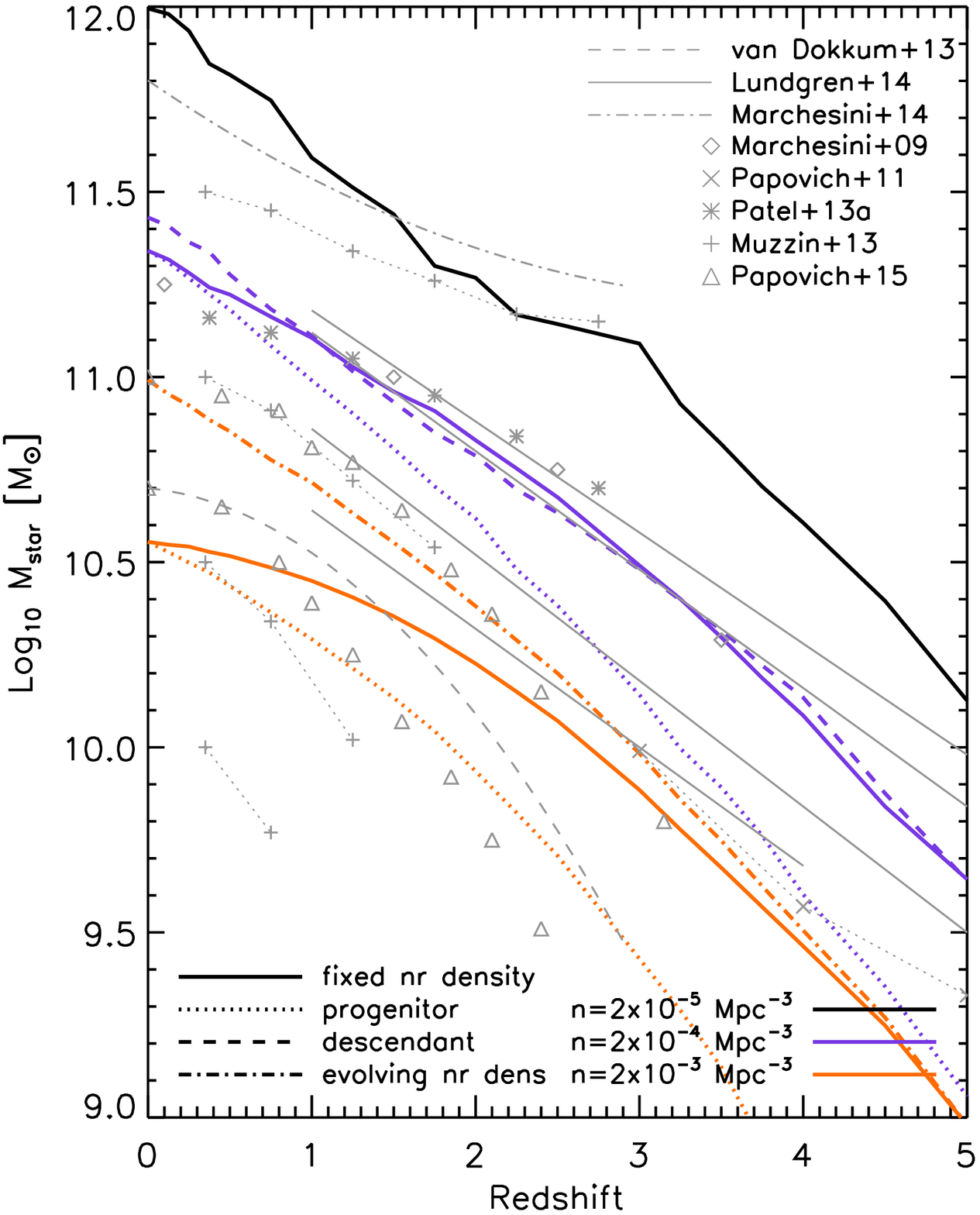}
\caption {\label{fig:data} Evolution of the median stellar mass using different methods to link high-redshift with low-redshift galaxies at several different number densities. \emph{The simulation data has been scaled up by 0.5~dex to account for underproducing stellar masses due to too efficient feedback.} The solid curves show median masses at fixed number density, $n(>M)=2\times10^{-5}$~Mpc$^{-3}$ (black), $n(>M)=2\times10^{-4}$~Mpc$^{-3}$ (purple), and $n(>M)=2\times10^{-3}$~Mpc$^{-3}$ (orange). Progenitors (from $z=0$) and descendants (from $z=5$) are shown by coloured dotted and dashed curves, respectively. The orange dot-dashed curve shows the evolution at an evolving number density to take into account mergers. Grey data points and grey curves show a variety of observational results from the literature \citep{Marchesini2009, Papovich2011, Patel2013a, Dokkum2013, Muzzin2013, Lundgren2014, Marchesini2014, Papovich2015} as indicated in the legend.}
\end{figure}

In Figure~\ref{fig:data} we show our results from Figures~\ref{fig:nrprogdesc}, \ref{fig:nrevol5141}, and~\ref{fig:nrs} \emph{scaled up by 0.5~dex to account for underestimating stellar masses due to too efficient feedback}, together with a representative selection of observational data from the literature \citep{Marchesini2009, Papovich2011, Patel2013a, Dokkum2013, Muzzin2013, Lundgren2014, Marchesini2014, Papovich2015}. The solid curves show median masses at fixed number density, $n(>M)=2\times10^{-5}$~Mpc$^{-3}$ (black), $n(>M)=2\times10^{-4}$~Mpc$^{-3}$ (purple), and $n(>M)=2\times10^{-3}$~Mpc$^{-3}$ (orange). Progenitors (from $z=0$) and descendants (from $z=5$) are shown by coloured dotted and dashed curves, respectively. The orange dot-dashed curve shows the evolution at an evolving number density to take into account mergers. We caution again that the simulations were not tuned to reproduce the stellar mass function. We therefore focus on evolutionary trends and not on the normalization and believe the trends with redshift to be robust. A more in-depth discussion of the observational comparison can be found in Section~\ref{sec:obs}. 

The lowest-redshift number densities used in the observations are $2\times10^{-4}$~Mpc$^{-3}$ for \citet{Marchesini2009} and \citet{Papovich2011}, $1.4\times10^{-4}$~Mpc$^{-3}$ for \citet{Patel2013a}, $1.1\times10^{-4}$~Mpc$^{-3}$ for \citet{Dokkum2013}, $1.8\times10^{-4}$, $3.2\times10^{-4}$, $5.6\times10^{-4}$, and $1\times10^{-3}$~Mpc$^{-3}$ for \citet{Lundgren2014}, $3\times10^{-6}$~Mpc$^{-3}$ for \citet{Marchesini2014}, and $4\times10^{-4}$ and $1.3\times10^{-3}$~Mpc$^{-3}$ for \citet{Papovich2015}. \citet{Muzzin2013} base their number density selection on stellar mass and $n(>10^{11.5}~M_\odot)=3\times10^{-5}$, $n(>10^{11}~M_\odot)=5.6\times10^{-4}$, and $n(>10^{10}~M_\odot)=4.4\times10^{-3}$~Mpc$^{-3}$.

The way the galaxies were selected varies for the observational data. Most of the data shown use a fixed comoving number density as in the simulation data, but \citet{Papovich2015} use an evolving number density and the triangles should therefore be compared to the dot-dashed curve for which the number density evolves (based on the merger rate as shown in Figure~\ref{fig:leftsame}) from $n(>M)=2\times10^{-3}$~Mpc$^{-3}$ at $z=5$ to $n(>M)=0.64\times10^{-3}$~Mpc$^{-3}$ at $z=0$. \citet{Marchesini2014} also use an evolving number density, but the grey, dot-dashed curve can be compared to the black, solid curve, because in our simulations the mass evolution at an evolving number density is very similar to that at a fixed number density at $n(>M)\le2\times10^{-4}$~Mpc$^{-3}$ as also shown in Figure~\ref{fig:nrevol514}. Our fiducial number density ($n(>M)=2\times10^{-4}$~Mpc$^{-3}$; purple curves) follow similar evolutionary trends as the observational data. The discrepancy is larger at higher stellar masses, where the simulations show faster stellar mass growth than the observations of \citet{Muzzin2013} and \citet{Marchesini2014}. Quenching is thus potentially not sufficiently efficient in galaxy clusters. The discrepancy is also larger at lower stellar masses, where the observations of \citet{Dokkum2013} and \citet{Muzzin2013} show a faster growth at $1<z<3$. This could indicate that the simulations suppress star formation too much in this mass regime. We checked that our simulation without AGN feedback shows a similar trend as the one with AGN feedback (but with $\sim0.2$~dex higher masses). \citet{Torrey2015} and \citet{Clauwens2016}, using simulations that approximately match the stellar mass function, also show slower evolution at $z\approx3-0$ for $M_\mathrm{star}\approx10^{10-10.7}$~M$_\odot$ at $z=0$ than inferred from the data, so it seems to be a generic feature of the current generation of hydrodynamical simulations.

\label{lastpage}

\end{document}